\documentclass{ieeeaccess}
\usepackage{cite}
\usepackage{amsmath,amssymb,amsfonts}
\usepackage{algorithmic}
\usepackage{graphicx}
\usepackage{textcomp}

\usepackage{amssymb}
\usepackage{array}
\usepackage{tabularx}
\usepackage{lipsum}
\usepackage{comment}
\usepackage{caption}

\usepackage{amsmath}
\usepackage[acronym,nopostdot,nonumberlist]{glossaries-extra}
\usepackage{glossary-inline}
\usepackage[table]{xcolor}
\usepackage{longtable}
\usepackage{array}
\usepackage{multirow}
\usepackage{booktabs}
\usepackage[capitalise]{cleveref}
\usepackage{microtype}
\usepackage[margin=1in]{geometry}
\RequirePackage[figuresleft]{rotating}
\usepackage{url}

\usepackage{rotating}

\usepackage{xcolor}
\usepackage[normalem]{ulem}  

\def\BibTeX{{\rm B\kern-.05em{\sc i\kern-.025em b}\kern-.08em
    T\kern-.1667em\lower.7ex\hbox{E}\kern-.125emX}}
\begin{document}
\history{Date of publication xxxx 00, 0000, date of current version xxxx 00, 0000.}
\doi{10.1109/ACCESS.2017.DOI}

\title{A Scalable Framework for Safety Assurance of Self-Driving Vehicles based on Assurance 2.0}

\author{\uppercase{Shufeng Chen}\authorrefmark{1}, \uppercase{Mariat James Elizebeth}\authorrefmark{1}, \uppercase{Robab Aghazadeh Chakherlou}\authorrefmark{1}, \uppercase{Xingyu Zhao}\authorrefmark{1}, \uppercase{Eric Barbier}\authorrefmark{2}, \uppercase{Siddartha Khastgir}\authorrefmark{1}, \uppercase{Paul Jennings}\authorrefmark{1}}
\address[1]{WMG, University of Warwick, 6 Lord Bhattacharyya Way, Coventry, CV4 7AL, United Kingdom}
\address[2]{Wayve Technologies Ltd., 230-238 York Way, London, N7 9AG, United Kingdom}


\markboth
{Author \headeretal: Preparation of Papers for IEEE TRANSACTIONS and JOURNALS}
{Author \headeretal: Preparation of Papers for IEEE TRANSACTIONS and JOURNALS}

\corresp{Corresponding author: Shufeng Chen (e-mail: Shufeng.Chen.1@warwick.ac.uk).}

\begin{abstract}
Assurance 2.0 is a modern framework developed to address the assurance challenges of increasingly complex, adaptive, and autonomous systems. Building on the traditional Claims-Argument-Evidence (CAE) model, it introduces reusable assurance theories and explicit counterarguments (defeaters) to enhance rigor, transparency, and adaptability. It supports continuous, incremental assurance, enabling innovation without compromising safety. However, limitations persist in confidence measurement, residual doubt management, automation support, and the practical handling of defeaters and confirmation bias. This paper presents \textcolor{black}{a set of decomposition frameworks to identify a complete set of safety arguments and measure their corresponding evidence.} Grounded in the Assurance 2.0 paradigm, the framework is instantiated through a structured template and employs a three-tiered decomposition strategy. \textcolor{black}{A case study regarding the application of the decomposition framework in the end-to-end (E2E) AI-based Self-Driving Vehicle (SDV) development is also presented in this paper.} At the top level, the SDV development is divided into three critical phases: Requirements Engineering (RE), Verification and Validation (VnV), and Post-Deployment (PD). Each phase is further decomposed according to its Product Development Lifecycle (PDLC). To ensure comprehensive coverage, each PDLC is analyzed using an adapted 5M1E model (Man, Machine, Method, Material, Measurement, and Environment). Originally developed for manufacturing quality control, the 5M1E model is reinterpreted and contextually mapped to the assurance domain. This enables a multi-dimensional decomposition that supports fine-grained traceability of safety claims, evidence, and potential defeaters.
\end{abstract}


\titlepgskip=-15pt

\maketitle

\section{Introduction}
\label{sec:introduction}
\PARstart{S}{afety} cases are structured, evidence-based arguments that a system is acceptably safe for a specific application and context. They are essential in safety-critical domains such as aerospace, automotive, maritime, railway, and healthcare, where regulatory compliance and public trust are paramount. Traditional safety assurance approaches often rely on prescriptive standards, which can result in extensive documentation with limited adaptability or traceability. To address these limitations, goal-based frameworks such as the Goal Structuring Notation (GSN) have been developed. GSN provides a graphical method for articulating safety goals, strategies, and supporting evidence, and has become widely adopted due to its clarity and modularity \cite{eurocontrol2006safety}. Another foundational model is the Claims-Argument-Evidence (CAE) framework, which emphasizes logical structure and traceability in assurance cases \cite{adelardResources}\cite{CAE2}.

However, as systems become increasingly complex, adaptive, and autonomous, particularly with the integration of AI and Machine Learning (ML), traditional assurance methods struggle to scale and adapt \cite{bloomfield_disruptive_2019,zhao2020safety}. \textcolor{black}{Furthermore, the traditional method is expensive from both time and money points of view and is considered as a drag for innovation \cite{bloomfieldAssurance2}}. Assurance 2.0 is a modern framework developed to address the assurance challenges posed by increasingly complex, adaptive, and autonomous systems. Building on the traditional CAE model, it introduces reusable assurance theories and explicit counterarguments (defeaters) to enhance rigour, transparency, and adaptability \cite{bloomfieldAssurance2}. \textcolor{black}{It enables ongoing, stepwise assurance, allowing for innovation while maintaining safety.}

\begin{figure}[tb]
    \centering
    \includegraphics[width=0.95\linewidth]{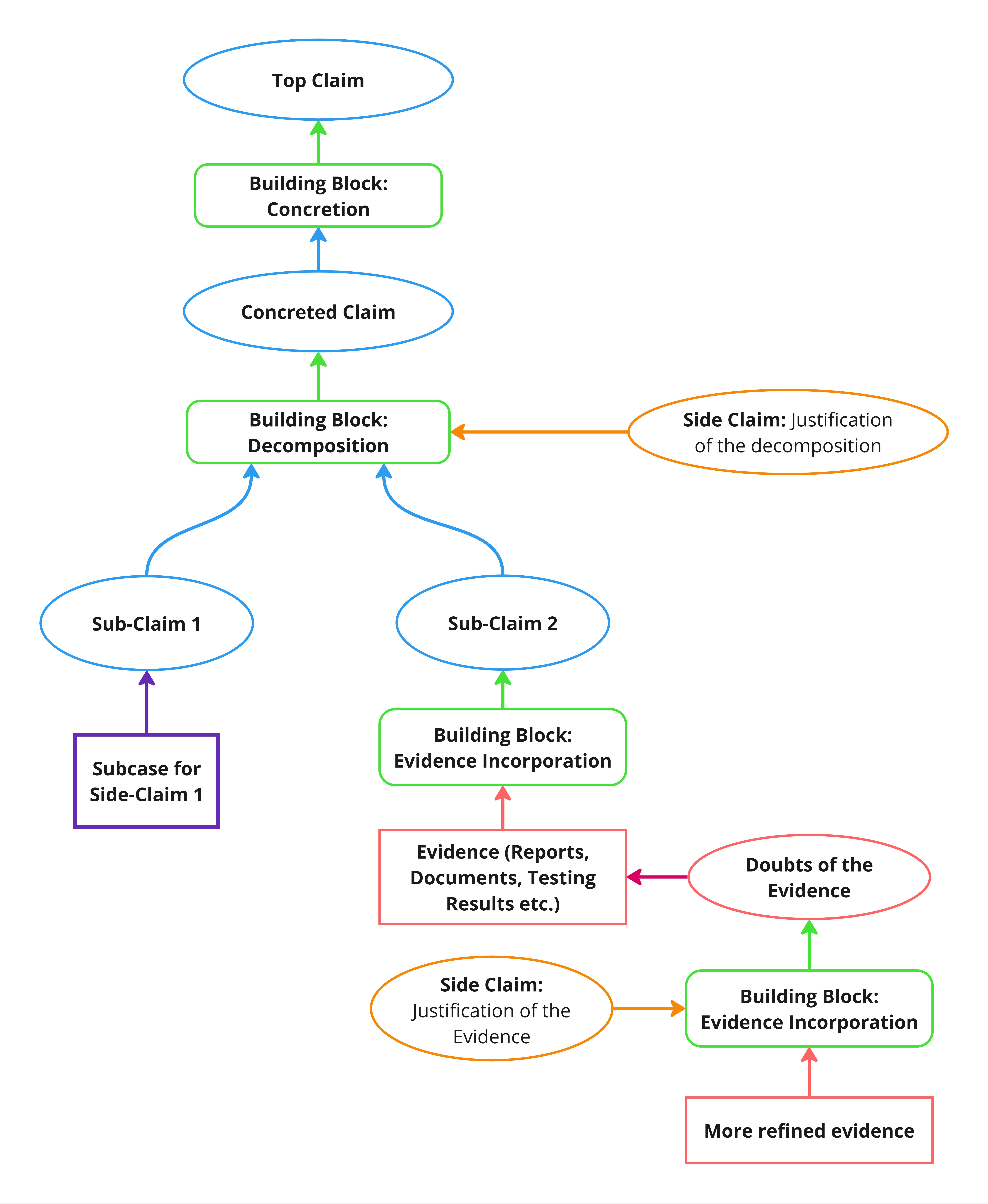}
    \caption{A Generic Representation of the Assurance 2.0 Diagram}
    \label{fig:generic diagram}
\end{figure}

A generic Assurance 2.0 diagram is presented in Figure \ref{fig:generic diagram}. On top of the \textbf{Claim} (or \textbf{Sub-claim}, \textbf{Sub-sub-claim}), \textbf{Argument}, and \textbf{Evidence} blocks that are already covered in the CAE model, Assurance 2.0 also introduces \textbf{Defeaters} representing challenges or counterarguments that must be addressed or explicitly accepted as residual risks. Assurance 2.0 also introduces \textbf{Side Claims}, represented as supporting assertions that help justify the reasoning steps in an assurance case. To effectively construct rigorous and logically sound assurance arguments, compared to the CAE model, the argument block in Assurance 2.0 can be categorized as five building blocks\cite{bloomfieldAssurance2}:
\begin{itemize}
    \item The \textbf{Concretion} block translates an abstract claim to a more specific one.
    \item The \textbf{Substitution} block replaces one claim with another that is logically equivalent or more useful.
    \item The \textbf{Decomposition} block breaks down a complex claim into simpler sub-claims.
    \item The \textbf{Calculation} block applies quantitative reasoning to support a claim.
    \item The \textbf{Evidence Incorporation} block introduces empirical or analytical evidence to support a claim.
\end{itemize}
These five building blocks ensure that every part of the argument contributes meaningfully to the overall assurance case.

\subsection{Research Gaps and Motivations}
With the increasing complexity of systems, especially those equipped with AI-driven decision-making systems, providing a structured framework to prove that the system is safe (or unsafe) \textcolor{black}{to an acceptable level} has become a formidable challenge. Unlike traditional systems, these technologies operate in vast, unpredictable environments and rely on probabilistic reasoning rather than deterministic logic. As a result, building a credible and comprehensive safety assurance case requires more than conventional methods. It demands a structured, evidence-driven approach that can handle complexity, uncertainty, and scale.

One of the core difficulties lies in the incompleteness of testing and analysis. The operational domain is too expansive to be exhaustively covered, and many safety arguments are inherently statistical. Moreover, the causal pathways to hazards in AI systems are often indirect and multifaceted, making it harder to identify all potential risks and ensure that safety requirements are fully addressed.

\textcolor{black}{Assurance 2.0 provides a framework for evaluating dependability features of the System of Interest (SoI), such as reliability, availability, and safety which is the focus of this paper.} Despite its conceptual strengths, Assurance 2.0 currently lacks a standardized decomposition framework \textcolor{black}{to ensure that all the relevant safety arguments are covered}. This absence limits the completeness of assurance cases, as it becomes difficult to ensure that all relevant safety concerns, evidence, and counterarguments are consistently captured, particularly in complex \textcolor{black}{systems such as SDV.}

In parallel, the increasing scale, dynamism, and iterative nature of modern system development demand assurance processes that are not only rigorous but also highly efficient. Manual construction and maintenance of assurance cases are time-consuming, resource-intensive, and prone to inconsistencies, especially when systems evolve rapidly. Without integrated automation capabilities, such as argument generation, evidence integration, and defeater analysis. Assurance 2.0 struggles to support continuous assurance in agile or DevOps environments.

Furthermore, the communication of safety rationale remains a challenge. As assurance cases grow in size and complexity, stakeholders from different disciplines often find it difficult to interpret the structure, trace the logic, or assess the residual risks. A lack of standardized decomposition and automation hinders the clarity, traceability, and accessibility of safety arguments, reducing their effectiveness as tools for cross-functional communication and regulatory engagement.

A set of structured safety case templates tailored for autonomous systems is presented in \cite{bloomfield2021safety}, particularly those incorporating machine learning (ML) components. The templates aim to support the development, deployment, and evolution of such systems by offering reusable argument structures for key assurance activities. They also support handling of defeaters (i.e., counterarguments) and evidence confidence, aligning with the Assurance 2.0 paradigm. While the templates offer a valuable foundation for structuring safety arguments, several limitations affect their ability to ensure completeness and usability:

\begin{itemize}
    
    \item \textbf{Lack of Formal Decomposition Framework:} There is no underlying universal decomposition model (e.g., based on lifecycle stages or operational dimensions like 5M1E) to guide systematic and exhaustive argument development.
    
    \item \textbf{Limited Automation Support:} The templates are static and manually applied, lacking integration with tools that could automate argument generation, evidence linking, or defeater analysis.
    
    \item \textbf{Scalability Challenges:} As system complexity increases, manually adapting and maintaining these templates becomes increasingly difficult, potentially leading to gaps in coverage or inconsistencies in argumentation.
\end{itemize}

Therefore, to tackle the aforementioned limitations, the authors \textcolor{black}{developed a tailored template to cover safety arguments of all system lifecycles}, with the aim of addressing the following research questions: 

\begin{itemize}
    \item \textbf{RQ1 (Completeness of a Safety Case):} How can a universal decomposition framework ensure comprehensive coverage of safety arguments and evidence across system lifecycles and domains?
    
    \item \textbf{RQ2 (Efficiency of a Safety Case Development):} How can automation improve the efficiency of developing, maintaining, and validating Assurance 2.0-based safety cases?
    
    \item \textbf{RQ3 (Communication of a Safety Case):} How can structured and automated assurance frameworks enhance the clarity, traceability, and stakeholder understanding of safety cases?
\end{itemize}

These challenges motivate the need for a universal, automation-ready decomposition framework that enhances the completeness, efficiency, and communicability of Assurance 2.0-based safety cases.
\subsection{Paper Contributions}
When developing the safety case of a complex system, several key questions must be addressed: 

\textbf{(i)} Do we understand, and have we clearly described what is being developed? 

\textbf{(ii)} Do we appreciate the scale of the challenge and how it is being managed? 

Safety case templates support this process by guiding practitioners to better understand and document their system \cite{bloomfield2021safety}. The main contribution of this work is to provide a guideline that encompasses all factors and elements influencing the assurance case across the different phases of the lifecycle. This paper proposes a systematic decomposition strategy that transforms high-level safety claims into well-structured, traceable sub-claims. The proposed approach is designed to ensure:

\begin{itemize}
    \item \textbf{Completeness:} All relevant aspects of safety are systematically addressed based on the decomposition frameworks \textcolor{black}{inspired from well-recognised frameworks like the V-Model, and 5M1E \cite{ISO9001}. This helps consider all relevant system development phases and factors of each phase. Primarily focused on SDV applications, the template can be tailored for applications in other domains}.
    \item \textbf{Efficiency:} Drawing the diagrams in parallel with the development can be time-consuming, bringing additional challenges when changes are required. A set of Excel-based templates, based on the decomposition frameworks, is created to support the documentation. These templates can then be used as the inputs for the automation toolchain to develop the final diagrams.
    \item \textbf{Clarity and Traceability:} Based on the decomposition frameworks, each claim is logically organized and mutually exclusive, and is explicitly supported by verifiable evidence.
\end{itemize}

At the core of this strategy are decomposition frameworks, which are structured reference models derived from industry standards, regulatory guidance, and engineering best practices. These frameworks provide a disciplined foundation for constructing assurance cases that are rigorous, scalable, and auditable.

\textcolor{black}{The remaining part of the paper is organised as follows: Section II provides a literature review of relevant study in the field of safety case, Assurance 2.0, relevant standards, AI system development and its safety assurance, and the 5M1E model; Section III introduces the proposed methodology in this paper; Section IV presents the case study of the proposed framework in SDV; Section V summarizes the discussion of the results and the framework; and Section VI concludes our work.}

\section{Literature Review}
\subsection{Preliminary}
\subsubsection{Safety Case}
A safety case is a structured, evidence-based argument that a system is acceptably safe for its intended use in a specific operational context. It serves as a cornerstone for regulatory approval, stakeholder confidence, and lifecycle safety assurance. Traditionally, safety cases have been manually constructed and evaluated, often as a final step in the development process. However, this approach is increasingly viewed as insufficient for modern, complex systems that evolve rapidly and operate in dynamic environments. As a result, the field has seen a shift toward model-based and incremental safety case methodologies, which integrate safety assurance throughout the system lifecycle.


Goal Structuring Notation (GSN) is a widely used graphical approach for organizing safety cases. It visually represents the logical flow of safety arguments by breaking down high-level goals into sub-goals, strategies, and supporting evidence \cite{GSN}\cite{GSN1}. Its structured format can help stakeholders identify potential gaps in reasoning or missing evidence. However, as noted in \cite{GSNNancy}, while GSN is effective for presenting safety arguments, it does not directly address the underlying assurance process. Specifically, it emphasizes how arguments are structured rather than how their content is validated, and it offers limited guidance on assessing the quality or sufficiency of supporting evidence. In practice, GSN is sometimes applied retrospectively to document decisions already made, rather than serving as a proactive tool for guiding assurance activities. These observations are not unique to GSN and can also apply to other notational frameworks like CAE, highlighting the importance of focusing on the assurance methodology itself rather than the notation used to represent it.

\subsubsection{Assurance 2.0}
In addition to GSN, the Assurance 2.0 framework has emerged as a state-of-the-art alternative. Developed by Bloomfield and Rushby, Assurance 2.0 extends traditional safety case structures by incorporating defeaters, confirmation theory, and parameterized assurance theories \cite{bloomfield2024nutshell}. It emphasizes eliminative argumentation, encouraging developers to actively seek and refute potential safety concerns rather than merely confirm safety. Assurance 2.0 also supports probabilistic reasoning and dialectical examination, aiming for indefeasible confidence, which is a state where no credible new information would overturn the safety judgement \cite{bloomfield2024confidence}.

\subsubsection{Relevant Standards}
The automotive industry is governed by a suite of safety standards designed to ensure that vehicles meet rigorous safety requirements throughout their lifecycle. These standards form the backbone of safety case development, providing structured methodologies for hazard analysis, risk mitigation, and verification. As mentioned in \cite{bloomfield_disruptive_2019}, in the case of E2E AI System, it is difficult to rely solely on a standards-based justification given the lack of validated standards, policies, and guidance for such novel technologies. 

ISO 26262 is the cornerstone of functional safety in automotive systems \cite{ISO26262}. It is derived from the general safety standard IEC 61508 and tailored specifically for electrical and electronic systems in road vehicles \cite{IEC61508}. The standard defines a V-model-based development lifecycle, encompassing concept, design, implementation, verification, validation, and decommissioning.

Within the V-model, requirement engineering is a critical part that sits in parallel with the concept and design phase. ISO/IEC/IEEE 29148:2018 defines the processes and best practices for requirement engineering across systems and software lifecycles \cite{ISO29119}. It provides guidance on eliciting, analysing, specifying, and managing requirements, which is crucial for safety-critical automotive systems.

ISO/IEC/IEEE 29119 is a multi-part standard that defines a comprehensive framework for software testing \cite{ISO29119}. It includes processes, documentation templates, and techniques applicable across industries. ISO/IEC/IEEE 29119 can be referred to as the recommended practices for the other side of the V-model related to verification and validation.

While ISO 26262 provides a robust framework for managing risks due to system failures, it does not address hazards arising from functional insufficiencies or performance limitations, which is a gap that becomes critical in autonomous and advanced driver-assistance systems (ADAS). To complement ISO 26262, ISO/PAS 21448 (SOTIF) addresses risks not caused by system malfunctions, but by limitations in perception, decision-making, or specification \cite{ISO21448}. This is particularly relevant for systems relying on sensors and AI, where the environment is unpredictable and the driving task is not fully specifiable.

ISO/TS 5083 is a technical specification designed to provide a holistic safety framework for automated driving systems (ADS), particularly those at SAE Levels 3 and 4 \cite{ISO5083}. It serves as an overarching standard that integrates and builds upon existing safety standards to address the unique challenges posed by automation, AI, and the interactions of complex systems. ISO/TS 5083 builds on ISO 26262 by extending its principles to automated driving contexts,  where the driver is no longer the fallback safety mechanism. It complements SOTIF by providing a structured safety framework for ADS that includes SOTIF principles. It also treats AI algorithms as software components requiring safety artifacts and validation steps, aligning with ISO/PAS 8800's guidance on AI lifecycle and risk management.

ISO 9001 is an internationally recognized standard for quality management systems (QMS) \cite{ISO9001}. It provides a framework for organizations to consistently deliver products and services that meet customer and regulatory requirements. In the context of safety-critical systems, ISO 9001 supports the systematic management of quality-related risks, complements standards like ISO 26262, and provides a foundation for traceability, documentation, and auditability. ISO 9001 also advocates the use of the 5M1E model, which is a structured approach used in root cause analysis and problem-solving. It categorizes potential causes of a problem into six domains, including Man (Human), Machine, Material, Method, Measurement, and Environment.

\subsubsection{Development Lifecycle for AI Systems}
As AI becomes increasingly embedded in safety-critical automotive systems, particularly in ADS and ADAS, traditional safety standards such as ISO 26262 and ISO 21448 (SOTIF) are no longer sufficient to address the unique risks posed by AI technologies. To fill this gap, ISO/PAS 8800:2024 was introduced as a dedicated framework for managing the safety of AI-based systems in road vehicles \cite{ISO8800}. ISO/PAS 8800 defines a structured AI safety lifecycle that aligns with and extends the V-model used in ISO 26262. This includes the identification of safety goals and hazards specific to AI behaviour, definition of system boundaries and interfaces for AI components, validation of training and testing data quality and their coverage, runtime monitoring of AI behaviour and performance, and continuous learning and adaptation based on real-world data.

\subsection{Related Work}
\subsubsection{Safety Assurance of Autonomous Systems}
The safety assurance of autonomous systems, particularly those incorporating ML, presents unique challenges that traditional safety frameworks struggle to address. Recent literature has proposed methodologies and frameworks that aim to address the challenges. 

Paterson et al. proposed the framework of AMLAS (Safety Assurance of Autonomous Systems) in \cite{AMLAS}. AMLAS is a methodology for integrating safety assurance into the ML development lifecycle for autonomous systems. It promotes structured safety case development using GSN. AMLAS emphasizes evidence generation across the ML lifecycle, from data collection to deployment. It also supports eliminative reasoning by identifying and mitigating performance limitations and data biases. However, there are also limitations associated with AMLAS framework in its current form. The first limitation relates to the complexity of evidence generation process. Constructing a compelling safety case for ML components requires novel types of evidence, such as data quality assessments, model robustness evaluations, and performance metrics under diverse conditions. AMLAS provides guidance, but the generation and validation of such evidence remain complex and resource-intensive. Secondly, AMLAS focuses primarily on development-time assurance. While it supports post-deployment monitoring, it does not fully address the challenges of runtime learning or adaptation, which are increasingly relevant in autonomous systems. Furthermore, the process of AMLAS is largely conceptual and relies on manual processes. There is limited tooling support for automating AMLAS activities such as evidence management and safety case generation.

Dong et al. introduces a quantitative framework for assessing the reliability of ML components in autonomous systems \cite{dong2023reliability}. It provides probabilistic safety arguments that quantify uncertainty and confidence. It also structures assurance cases with model-agnostic reliability metrics, enabling traceable and defensible claims. The framework supports the construction of indefeasible confidence by integrating statistical evidence and robustness verification into safety cases, which complements the current version of Assurance 2.0. While the proposed approach is innovative and practical, the authors acknowledge several limitations. The framework focuses primarily on design-time assurance. Firstly, while it supports robustness verification and offline testing, similar to AMLAS, it does not fully address runtime monitoring, adaptation, or online learning. Secondly, the safety argument templates and reliability models require manual instantiation and validation, which may not scale well for large systems with multiple ML components. The lack of automation tools could also hinder adoption in industrial settings. Lastly, although the paper proposes safety argument patterns, it does not fully explore how these ML-specific arguments integrate with system-level safety cases, especially those based on Assurance 2.0 or GSN frameworks.

Pai explores the conceptual and practical challenges of integrating ML components into safety-critical systems, particularly from the perspective of system safety engineering \cite{pai}. The author argues that ML introduces epistemic uncertainty due to its data-driven nature, which complicates traditional safety assurance approaches. The paper emphasizes safety as a system-level property and must be assessed in the context of the entire socio-technical system, instead of only assuring the ML model. The paper also advocates for closer integration between safety engineers and ML practitioners to develop assurance strategies that account for both performance and safety. However, the approach is largely theoretical and does not provide a concrete methodology or framework for implementing its ideas in practice. It also lacks case studies or experimental results to demonstrate how the proposed concepts can be applied to real-world systems. Lastly, while the paper discusses the limitations of traditional safety approaches, it does not explicitly map its ideas to standards like ISO 26262, ISO 21448 (SOTIF), or ISO/PAS 8800.

\subsubsection{Safety Assurance based on 5M1E Model}
\textcolor{black}{The 5M1E model is a structured framework commonly used in root cause analysis, quality management, and safety engineering to identify and categorize potential factors that contribute to a problem or risk. It helps ensure a comprehensive examination of all relevant aspects of a system or process. Originated in the manufacturing and quality management domain, 5M1E consists of six factors - i.e., Man (human factors), Machine (equipment-related issues), Material (raw material quality), Method (procedures and workflows), Measurement (accuracy and consistency of data and instruments), and Environment (physical conditions like temperature, humidity, lighting) \cite{5m1echemical}.} While the 5M1E model is not a formal safety case methodology, it is highly compatible with safety case development by supporting hazard identification and classification of causal factors, and structuring evidence for risk mitigation and quality control. Some existing literature demonstrates the potential of application in the early stages of safety case development.

Guo identified and categorized behavioral and systematic risk factors contributing to accidents in the construction industry based on the elements of the 5M1E model \cite{5m1eanalysis}. The model was used to structure a Bayesian network-enhanced risk matrix, enabling quantitative evaluation of unsafe behaviours and their impact on accident likelihood. While not a formal safety case, this application demonstrates how 5M1E can support hazard identification and argument structuring in safety-critical systems. In a chemical analysis laboratory for special equipment safety, the 5M1E model was used to build a QMS that ensures safe and accurate operation \cite{5m1echemical}. The model helped identify and control variables affecting test reliability, including human errors, equipment calibration and maintenance, and environmental and procedural consistency. This structured approach to quality aligns with ISO 9001 and can be extended to safety case evidence generation, particularly in demonstrating process control, traceability, and preventive measures.
\section{Methodology}
\textcolor{black}{To ensure the completeness of safety arguments for SDV,} in this paper, an Assurance 2.0 template integrated with a three-tiered decomposition model is introduced. \textcolor{black}{Starting with the vehicle-level top claim (i.e., vehicle is acceptably safe), the decomposition of the vehicle-level} safety arguments is then grounded in established systems engineering and quality management principles:

\begin{itemize}
    \item \textbf{Level 1: V-Model-Based Decomposition} \\
    Informed by the Systems Engineering Body of Knowledge (SEBoK) \cite{SEBoK2024}, \textcolor{black}{the Level-1 decomposition restructures the vehicle development into three phases, from Requirement Engineering (RE) Phase to Verification and Validation (VnV), and then Post-Deployment (PD) monitoring. This ensures alignment with engineering workflows and lifecycle traceability.} 

    \item \textbf{Level 2: Standards-Aligned Lifecycle Decomposition} \\
    Based on the decomposed phases from Level 1, each phase-level claim is then further decomposed into corresponding Product Development Lifecycles (PDLC) in the Level 2 decomposition. The set of decomposed PDLCs is guided by relevant international standards. It is important to note that the standards used in this template are example standards for SDV. They can vary based on the field under study.
    
    1) The RE phase is decomposed based on PDLCs suggested in ISO/IEC/IEEE 29148 \cite{ISO29148} and ISO 24766 \cite{ISO24766}.
    
    2) The VnV phase is decomposed based on PDLCs suggested in ISO/IEC/IEEE 29119 \cite{ISO29119} and ISO 26262 \cite{ISO26262}.
    
    3) The PD phase is decomposed based on the PDLCs suggested in ISO/TS 5083:2025 \cite{ISO5083}.

    \item \textbf{Level 3: 5M1E-Based Decomposition} \\
    \textcolor{black}{To ensure that all factors in each of the PDLC are considered, in Level 3 decomposition, each PDLC-level claim is further decomposed based on the 5M1E model.} Adapted from ISO 9001 quality management principles, this level categorizes influencing factors into six domains: \textit{Man}, \textit{Machine}, \textit{Method}, \textit{Material}, \textit{Measurement}, and \textit{Environment}. This enables a holistic analysis of safety risks across human, technical, and contextual dimensions.
\end{itemize}

By integrating these decomposition levels, the proposed framework enhances the transparency, defensibility, and adaptability of safety assurance cases, which are the set of criteria that need to be met for regulatory approval, stakeholder confidence, and long-term system reliability. This structured approach not only strengthens the credibility of safety arguments but also accelerates the safe and scalable deployment of autonomous technologies.

\subsection{Level 1 Decomposition of Safety Arguments based on the V-Model}

\begin{figure}[tb]
    \centering
    \includegraphics[width=\linewidth]{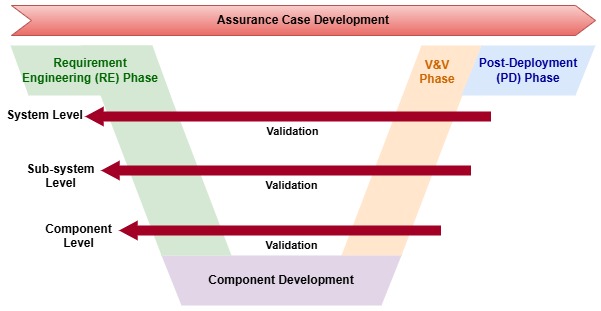}
    \caption{A Generic V-Model consisting of Requirement Engineering Phase, Verification and Validation Phase, Product Development Phase, and Post-Deployment Phase}
    \label{fig:V-Model}
\end{figure}

The V-Model is a product development framework used in systems engineering. It is widely used in safety-critical systems, such as aerospace \cite{ARP4761}, automotive \cite{ISO26262}, and medical devices \cite{ISO13485}. It represents the development process in a V-shaped diagram, emphasizing a correspondence between the development phase, the testing phase, and the post-deployment phase. The V-model promotes early test planning and parallel development and testing. A generic V-Model is illustrated in Figure \ref{fig:V-Model}. Horizontally, following the flow of the pipeline, product development begins with the Requirement Engineering (RE) Phase, which identifies product requirements at the system, subsystem, and component levels. Based on the requirements, the product was developed in the Component Development Phase. This is followed by the Verification and Validation (VnV) Phase at the system, sub-system, and component levels to ensure that the requirements are met. After the product has been fully verified, validated, and deployed to the end users or production environment, the system is monitored and maintained in the Post-Deployment (PD) Phase. Based on the monitoring results, the requirements or test plans may be updated to improve the system. 

\begin{figure}[tb]
    \centering
    \includegraphics[width=0.95 \linewidth]{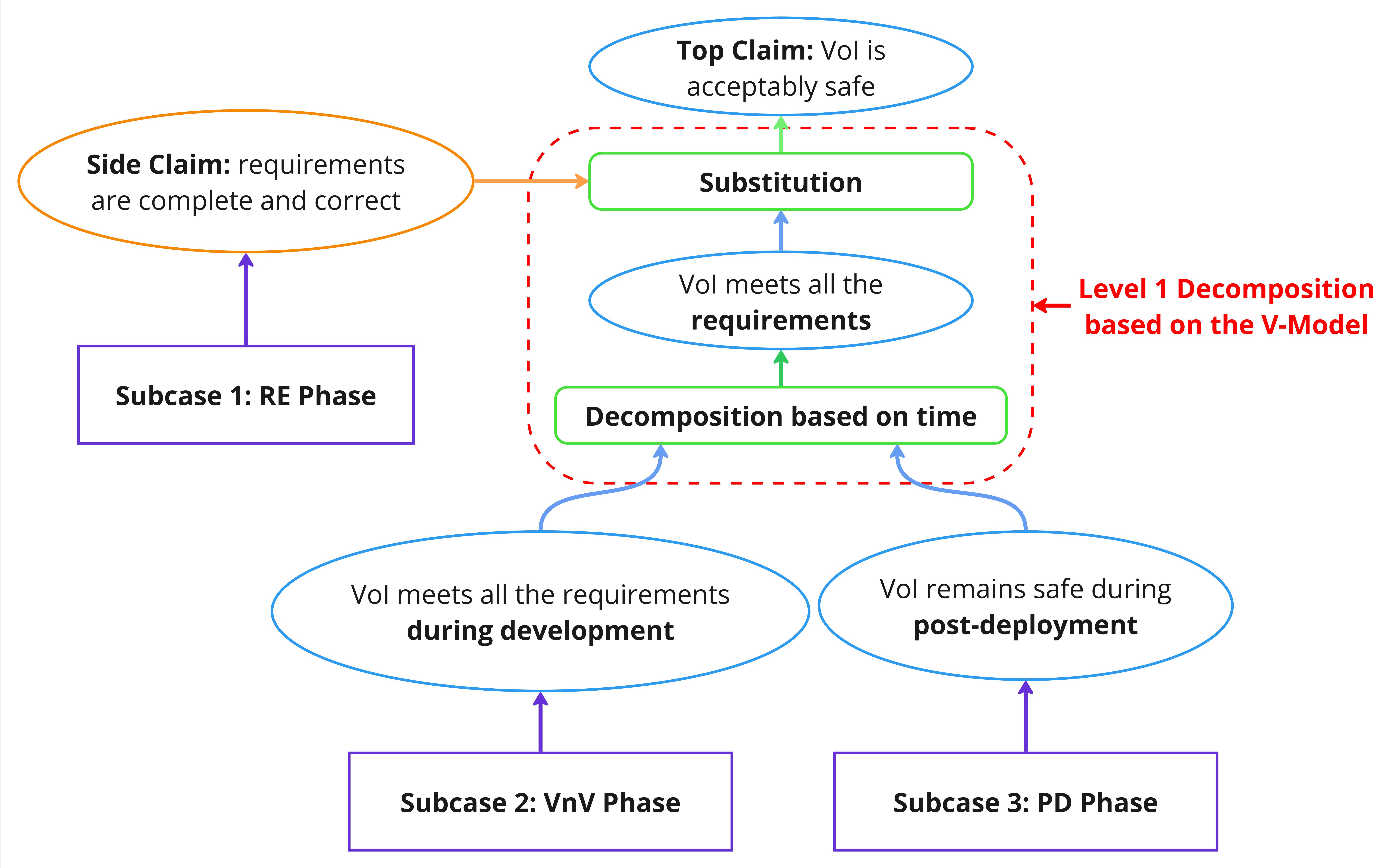}
    \caption{Level 1 Decomposition based on the V-Model and its Subcases}
    \label{fig:L1 Decom}
\end{figure}

In this work, the V-Model was applied as the Level 1 Decomposition framework for the vehicle-level claim, illustrated in Figure \ref{fig:L1 Decom}. The Level 1 decomposition follows the same logic as the safety case template proposed in \cite{adelardResources}. It starts with the top claim that the Vehicle of Interest (VoI) is acceptably safe. The top claim is then substituted using the \textbf{Substitution} block to a more tangible claim that the VoI meets all the requirements. A side claim is then added to justify that the requirements being used for testing the VoI are complete and correct. The decomposition at this level results in three subcases, including the subcases for RE, VnV, and PD phases. It is important to note that the Component Development Phase is not separately presented in the assurance case. Instead, it is embedded within the RE phase.

\subsection{Level 2 Decomposition of Safety Arguments based on the Lifecycles}
Once the three phases of the V-Model (i.e., the RE, VnV, and PD phases) have been identified, each phase is then further decomposed based on the PDLC within each phase. Due to the fundamentally different nature of the three phases, the decomposition frameworks applied at Level 2 are distinct for each phase. These tailored frameworks will be elaborated upon in the following subsections.

\subsubsection{Level 2 Decompositions of the Requirement Engineering Phase}
Requirements Engineering is a critical discipline within product development that focuses on identifying, analysing, documenting, validating, and managing stakeholder needs and system requirements. It defines what a system should do and under what constraints, serving as the foundation for successful system (and subsystems) development.

In this work, the PDLC outlined in ISO/IEC/IEEE 29148 (Requirement Engineering) and ISO/IEC 24766 (Systems and software engineering — Guide for requirements engineering tool capabilities) are adopted as the Level 2 decomposition framework for the RE phase. ISO/IEC/IEEE 29148 is an international standard that specifies best practices and processes for requirements engineering in systems and software engineering. ISO/IEC 24766 provides guidance on applying software life cycle processes specifically to safety-critical systems, ensuring that safety considerations are integrated throughout the development process. According to these standards, the requirements engineering process comprises the following key lifecycle stages:
\begin{itemize}
    \item Requirements Elicitation
    \item Requirements Analysis
    \item Requirements Specification
    \item Requirements Checking
    \item Requirements Management
\end{itemize}
Table \ref{Tab:PDLC RE} summarizes the objectives and key activities associated with each of these life cycle stages within the RE phase.

\begin{table}
\caption{PDLC of Requirement Engineering Phase}
\label{Tab:PDLC RE}
\centering
\begin{tabular}{|>{\centering\arraybackslash}p{0.1\textwidth}|p{0.3\textwidth}|} \hline 
\textbf{PDLC of RE Phase} & \textbf{Goals and Activities}\\
\hline
\textbf{Requirement Elicitation} & 
To understand the needs for the product and gather high-level requirements from stakeholders. Activities include:

\begin{itemize}
    \item Interview;
    \item Surveys;
    \item Workshops;
    \item Observation with stakeholders, including customers, engineers, and regulatory bodies;
\end{itemize}
\\
\hline

\textbf{Requirement Analysis} &
To refine, prioritize, and resolve conflicts in the gathered requirements. Activities include:
\begin{itemize}
    \item Analyse requirements for completeness, consistency, and feasibility;
    \item Prioritize requirements and resolve conflicts;
\end{itemize}
\\
\hline

\textbf{Requirement Specification} & 
To document the formal, testable, and unambiguous requirements in a structured, clear, and traceable format. Activities include:
\begin{itemize}
    \item Create detailed requirement specifications, use cases, and user stories;
    \item Ensure traceability; 
\end{itemize}
\\
\hline

\textbf{Requirement Checking} & 
To ensure that the requirements meet stakeholder needs and are feasible for implementation. Activities include:
\begin{itemize}
    \item Review requirements with stakeholders.
    \item Conduct simulations on the prototypes to validate requirements;
\end{itemize}
\\
\hline

\textbf{Requirement Management} &
To track changes, maintain traceability, and manage versions of the requirements. Activities include:
\begin{itemize}
    \item Maintain requirement traceability matrices;
    \item Manage requirement changes;
    \item Ensure alignment with project goals;
\end{itemize}
\\
\hline
\end{tabular}
\end{table}

\subsubsection{Level 2 Decompositions of the Verification and Validation Phase}
Verification and Validation are complementary processes used to ensure that a system or product meets its requirements and fulfils its intended purpose. Verification ensures that the product is being developed according to its specifications and design documents. Validation ensures that the final product meets the needs and expectations of stakeholders.
In this work, the VnV life cycles outlined in ISO/IEC/IEEE 29119 (Software Testing) are adopted as the Level 2 decomposition framework. ISO/IEC/IEEE 29119 is an international standard for software testing, providing a comprehensive framework that covers terminology, processes, documentation, techniques, and assessment models. It is designed to be applicable across all types of software development life cycles, including agile, waterfall, and V-models. According to the standard, the VnV process comprises the following key life cycle stages:
\begin{itemize}
    \item Test Requirement Analysis
    \item Test Planning
    \item Test Case Design
    \item Test Environment Setup
    \item Test Execution
    \item Test Cycle Closure
\end{itemize}
Table \ref{Tab:PDLC VnV} summarizes the objectives and key activities associated with each of these life cycle stages within the VnV phase.

\begin{table}
\caption{PDLC of Verification and Validation Phase}
\label{Tab:PDLC VnV}
\centering
\begin{tabular}{|>{\centering\arraybackslash}p{0.1\textwidth}|p{0.3\textwidth}|} \hline 
\textbf{PDLC of VnV Phase} & \textbf{Goals and Activities}\\
\hline
\textbf{Test Requirement Analysis} & 
To determine the required procedures for testing. Activities include:
\begin{itemize}
    \item Analyse requirements from a testing perspective;
    \item Identify testable and non-testable requirements;
    \item Determine types of testing needed (functional, performance, etc.);
    \item Define test coverage criteria;
\end{itemize}
\\
\hline

\textbf{Test Planning} &
To define the strategy and scope for testing. Activities include:
\begin{itemize}
    \item Test schedule (when to conduct the testing) and risk analysis;
    \item Estimate effort and resources;
    \item Update the estimations based on the post-deployment monitoring;
    \item Define roles and responsibilities;
    \item Allocate to the relevant test environment;
\end{itemize}
\\
\hline

\textbf{Test Case Design} &
To create detailed test cases. Activities include:
\begin{itemize}
    \item Create test scenarios and determine p/f criteria based on the system requirements;
    \item Create test scenarios and determine p/f criteria based on the post-deployment monitoring;
\end{itemize}
\\
\hline

\textbf{Test Environment Setup} &
To prepare the HW and SW conditions for testing. Activities include:
\begin{itemize}
    \item Set up servers, databases, and test tools;
    \item Configure test environments;
    \item Validate the environment with smoke testing (Smoke testing is a type of software testing that checks whether the most crucial functions of a program work properly after a new build or update. It’s often referred to as a "build verification test.")
\end{itemize}
\\
\hline

\textbf{Test Execution} &
To run the test cases and log results. Activities include:
\begin{itemize}
    \item Log defects in a defect tracking system;
    \item For each logged defect, retest or conduct a regression test as needed;
\end{itemize}
\\
\hline

\textbf{Test Cycle Closure} & To evaluate the testing process and outcomes. Activities include:
\begin{itemize}
    \item Analyse test results and metrics;
    \item Document lessons learned;
    \item Archive test artifacts (Archiving test artifacts during the Test Cycle Closure phase is the process of securely storing all relevant testing documents and outputs for future reference, compliance, audits, or knowledge transfer.);
\end{itemize}
\\
\hline
\end{tabular}
\end{table}

\subsubsection{Level 2 Decompositions of the Post-Deployment Phase}
The Post-Deployment Phase of the CAV system is a critical stage in the system’s lifecycle. It starts once the vehicle is released into real-world operation and continues throughout its service life. Since real-world conditions are far more complex than test environments, the post-deployment monitoring ensures that safety is maintained under all conditions. It also demonstrates that the system can improve over time to adapt to real-world conditions to build public confidence in CAV technology. Based on the results of the monitoring in the PD phase, the system requirements and test cases for system testing are also updated, enabling a feedback loop that drives innovation and refinement of AI models, control systems, and user experience.

In this work, the PDLCs outlined in ISO/TS 5083:2025 are adopted as the Level 2 decomposition framework for the PD phase. ISO/TS 5083:2025 is a technical specification that provides a comprehensive framework for ensuring the safety of automated driving systems in road vehicles. It focuses on Level 3 and Level 4 automation, as defined by ISO/SAE PAS 22736 \cite{ISO22736}, and covers the entire lifecycle of ADS – from design and development to the post-deployment phase. The PD phase is a key focus of ISO/TS 5083 because it recognizes that safety assurance does not end at launch. Instead, it must continue as the system operates in dynamic, real-world conditions. Based on the standard, the following key life cycle stages are considered in this work:
\begin{itemize}
    \item Operational Monitoring
    \item Incident and Event Handling (If it occurred)
    \item Change Management
    \item Field Update and Maintenance
\end{itemize}

Table \ref{Tab:PDLC PD} summarizes the objectives and key activities associated with each of these life cycle stages within the PD phase. It is important to note that in this work, the life cycle of Operational Monitoring involves both the field monitoring and cybersecurity monitoring, which are considered as separate parts in the ISO/TS 5083:2025.

\begin{table}
\caption{PDLC of Post-Deployment Phase}
\label{Tab:PDLC PD}
\centering
\begin{tabular}{|>{\centering\arraybackslash}p{0.1\textwidth}|p{0.3\textwidth}|} \hline 
\textbf{PDLC of PD Phase} & \textbf{Goals and Activities}\\
\hline
\textbf{Operational Monitoring} & To ensure the ADS continues to operate safely and securely, and as intended in real-world conditions. Activities include:
\begin{itemize}
    \item Collect real-time data from vehicles (e.g., via sensor logs, system states).
    \item Monitor for anomalies, edge cases, and unexpected behaviours.
    \item Monitor for vulnerabilities and intrusion attempts.
\end{itemize}
\\
\hline
\textbf{Incident and Event Handling}

\textbf{(If it occurred)} & To respond effectively to safety-critical events and prevent recurrence. Activities include:
\begin{itemize}
    \item Log and classify incidents (e.g., disengagements, near-misses).
    \item Perform incident analysis to identify the causes.
    \item Implement corrective actions and update safety documentation.
\end{itemize}
\\\hline
\textbf{Change Management} & To identify, assess, implement, and validate changes to the ADS once it is deployed. Activities include:
\begin{itemize}
    \item Impact assessment of system safety, performance, and compliance.
    \item Conduct regression testing to validate updates.
    \item System re-verification and re-validation.
    \item Configuration management.
    \item Safety case update.
    \item Stakeholder communication.
\end{itemize}
\\\hline
\textbf{Field Update and Maintenance} & To maintain and improve system performance without compromising safety. Activities include:

\begin{itemize}
    \item Deploy secure over-the-air (OTA) software updates.
    \item Schedule hardware inspections or recalls if needed.
\end{itemize}
\\\hline
\end{tabular}
\end{table}

\begin{figure}[tb]
    \centering
    \includegraphics[width=0.99 \linewidth]{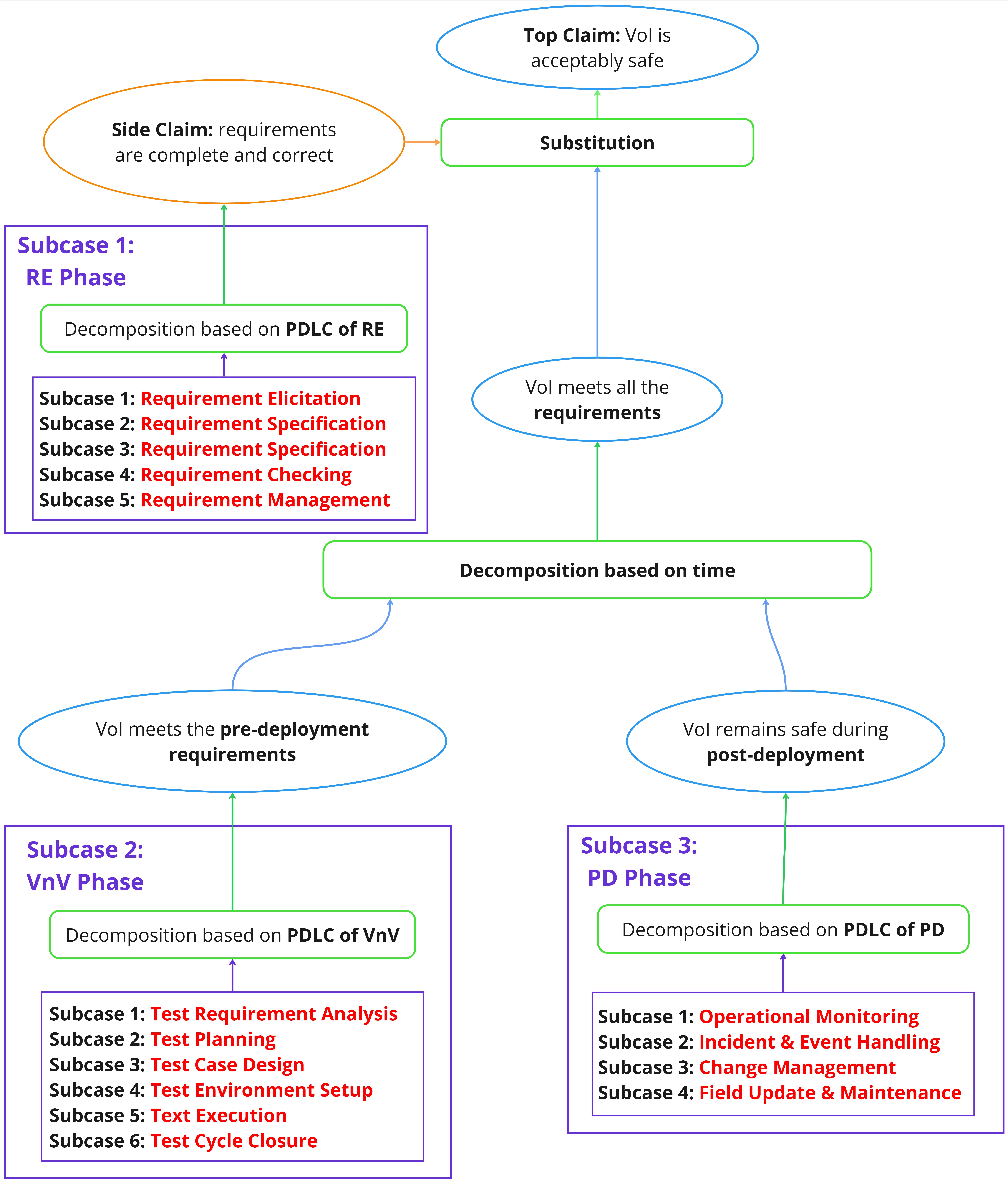}
    \caption{Level 2 Decompositions based on the PDLC and their Subcases}
    \label{fig:L2 Decom}
\end{figure}

Figure \ref{fig:L2 Decom} illustrates the proposed decomposition of each phase and the decomposed subcases in the Assurance 2.0 tree diagram.

\subsection{Level 3 Decomposition of Safety Arguments based on the 5M1E Model}
The 5M1E model is a structured framework for problem-solving and root cause analysis, widely used in quality management, particularly within manufacturing and industrial domains. It assists stakeholders in systematically identifying and categorizing potential causes of a problem, facilitating the development of effective solutions. Each letter in 5M1E represents a distinct category of potential causes. Table \ref{Tab:Old 5M1E} presents the traditional definitions of each factor within the manufacturing context, along with illustrative examples.

\begin{table}
\caption{Traditional Definitions of the 5M1E Model}
\label{Tab:Old 5M1E}
\centering
\begin{tabular}{|>{\centering\arraybackslash}p{0.1\textwidth}|p{0.3\textwidth}|} \hline
\textbf{Factors} & \textbf{Definitions} \\
\hline
\textbf{Man (Human)} & Human factors such as skills, training, behavior, or staffing. \\
\hline
\textbf{Machine} & Equipment, tools, or technology used in the process. \\
\hline
\textbf{Material} & Raw materials or components used in production. \\
\hline
\textbf{Method} & Processes, procedures, or techniques used. \\
\hline
\textbf{Measurement} & Accuracy and reliability of data collection and inspection. \\
\hline
\textbf{Environment} & External conditions that can affect the process. \\
\hline
\end{tabular}
\end{table}

\begin{figure}[tb]
    \centering
    \includegraphics[width=\linewidth]{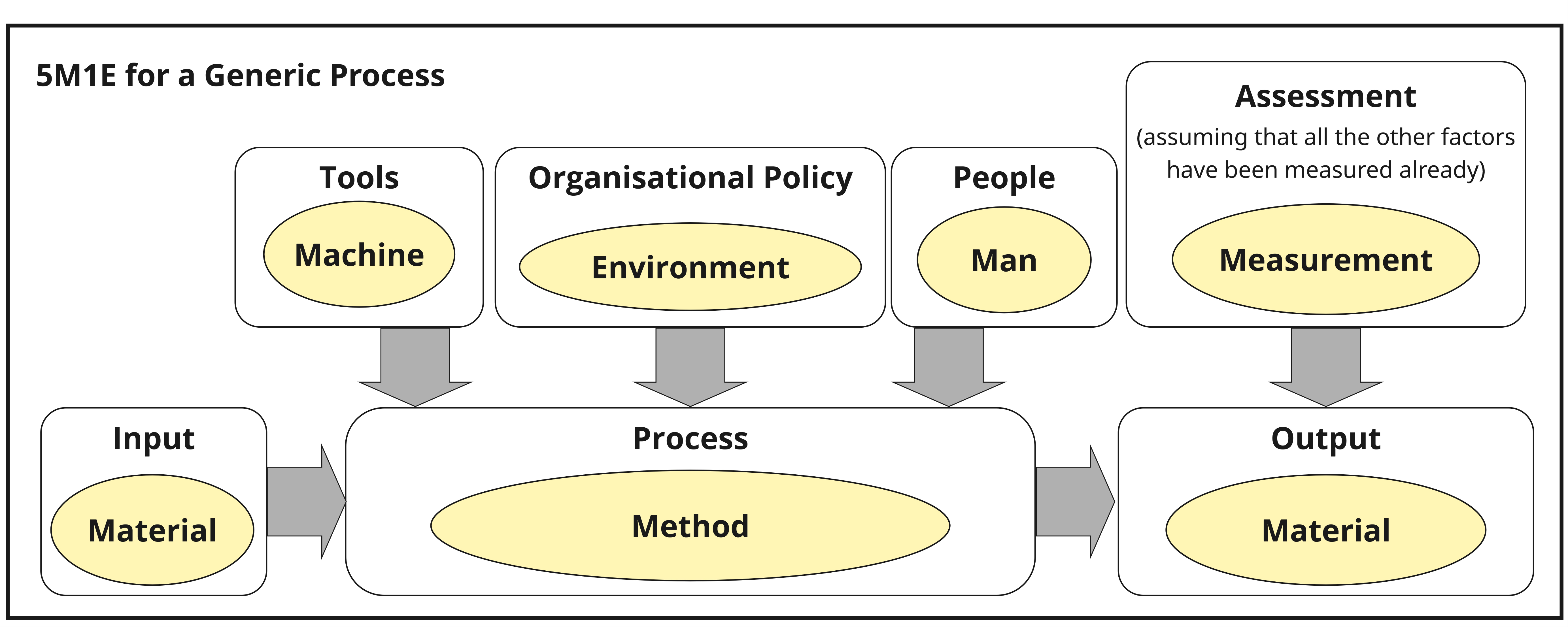}
    \caption{A Visualisation of the 5M1E of a Generic Process}
    \label{fig:Completeness of 5M1E}
\end{figure}

Based on the Assurance 2.0’s principle of natural language deductivism (NLD), we conceptually model a generic process as illustrated in Figure \ref{fig:Completeness of 5M1E}. In this NLD that we adopt, a process is described as follows: a person (Man), using Machines, governed by the organisational policy (Environment), applies a process (Method) to transform input Materials into the target output Materials. The effectiveness of the process can be directly assessed by measuring the outputs (Measurement). Note, although one could also measure inputs, machinery, or personnel, those measurements are simply outputs of other processes—e.g., evaluating software testers is itself the output of a recruitment process—thus preserving a consistent, layered assurance perspective.

\begin{figure}[tb]
    \centering
    \includegraphics[width=0.95 \linewidth]{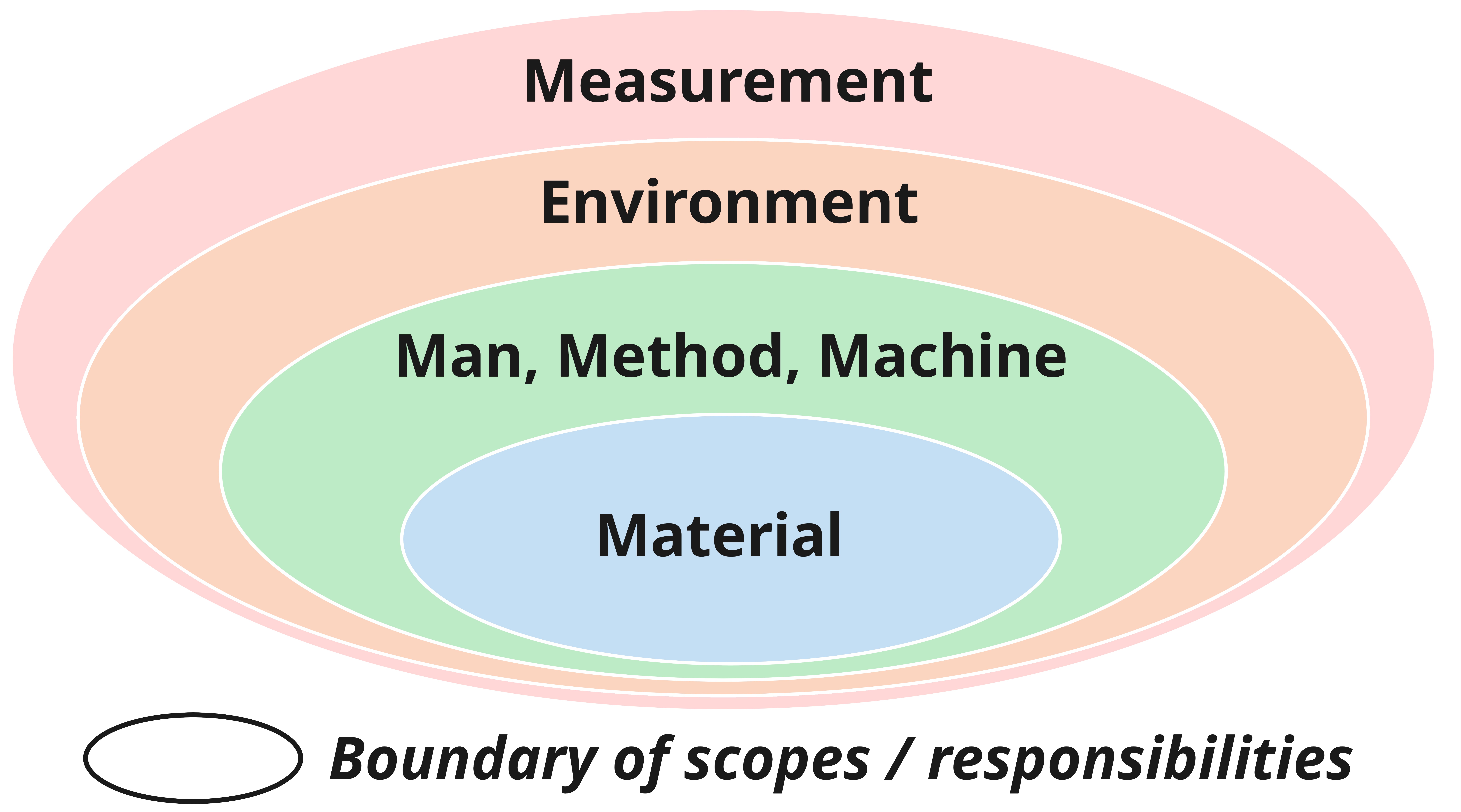}
    \caption{An Onion Diagram of the 5M1E Factors demonstrating their Interconnections based on the New Definition}
    \label{fig:Onion Diagram}
\end{figure}

The 5M1E model was selected as the level 3 decomposition framework in this work because it provides an exhaustive set of elements that need to be considered and assured in the safety case. However, since the 5M1E model was originally developed for manufacturing contexts, its direct application to structuring the assurance case for SDV requires adaptation. Specifically, the definitions of each of the six factors must be revised to reflect the unique characteristics and complexities of SDV development. Given the significantly greater complexity of SDV systems compared to traditional manufacturing processes, each factor in the 5M1E model must encompass a broader and more nuanced set of elements. To effectively support assurance case structuring, the factors must be redefined in a way that ensures they are mutually exclusive yet interconnected, allowing for comprehensive and non-overlapping analysis. Figure \ref{fig:Onion Diagram} presents an onion diagram illustrating the updated 5M1E structure. Each concentric ring represents the scope of a factor, with the Measurement factor forming the outermost layer. This positioning reflects its expanded role: rather than being limited to data collection (as in the original definition in Table \ref{Tab:Old 5M1E}), Measurement now encompasses the evaluation and monitoring of all other factors—Man, Machine, Material, Method, and Environment. Table \ref{Tab:New 5M1E} summarizes the revised definitions tailored for complex systems like SDVs. In the original model, measurements related to Man, Machine, Method, and Environment were embedded within those respective categories, limiting the model’s utility for safety assurance. By elevating Measurement to a distinct and overarching factor, the updated model provides a more structured and scalable framework for developing safety cases in complex, dynamic systems.

\begin{table}
\caption{Updated Definitions of the 5M1E Model for the Safety Assurance Framework}
\label{Tab:New 5M1E}
\centering
\begin{tabular}{|>{\centering\arraybackslash}p{0.1\textwidth}|p{0.3\textwidth}|} \hline
\textbf{Factors} & \textbf{Definitions} \\
\hline
Material (Core Layer)& The input, intermediate, and output \textbf{artefacts}.

Example: datasets, requirements, test results.\\
\hline
Method (Inner Layer)& The \textbf{processes} or \textbf{techniques }used to transform \textbf{Material}.\\
\hline
Machine (Inner Layer)& The \textbf{tools} or \textbf{technologies} used to support the creation of \textbf{Material}.\\
\hline
Man/Human
(Inner Layer)& The \textbf{roles} or \textbf{personnel} responsible for creating the \textbf{Material}.\\
\hline
Environment (Outer Layer)& The organizational context, policies, and physical conditions that enforce the \textbf{Method}, maintain the \textbf{Machine}, support \textbf{Man}, and protect \textbf{Material}.\\
\hline
Measurement (Outermost Layer)& The evaluation and monitoring of all other factors, ensuring that each component (\textbf{Man}, \textbf{Machine}, \textbf{Method}, \textbf{Material}, and \textbf{Environment}) is assessed for \textbf{quality}, \textbf{reliability}, and \textbf{safety}.\\
\hline
\end{tabular}
\end{table}

\begin{figure*}[tb]
    \centering
    \includegraphics[width=0.95 \linewidth]{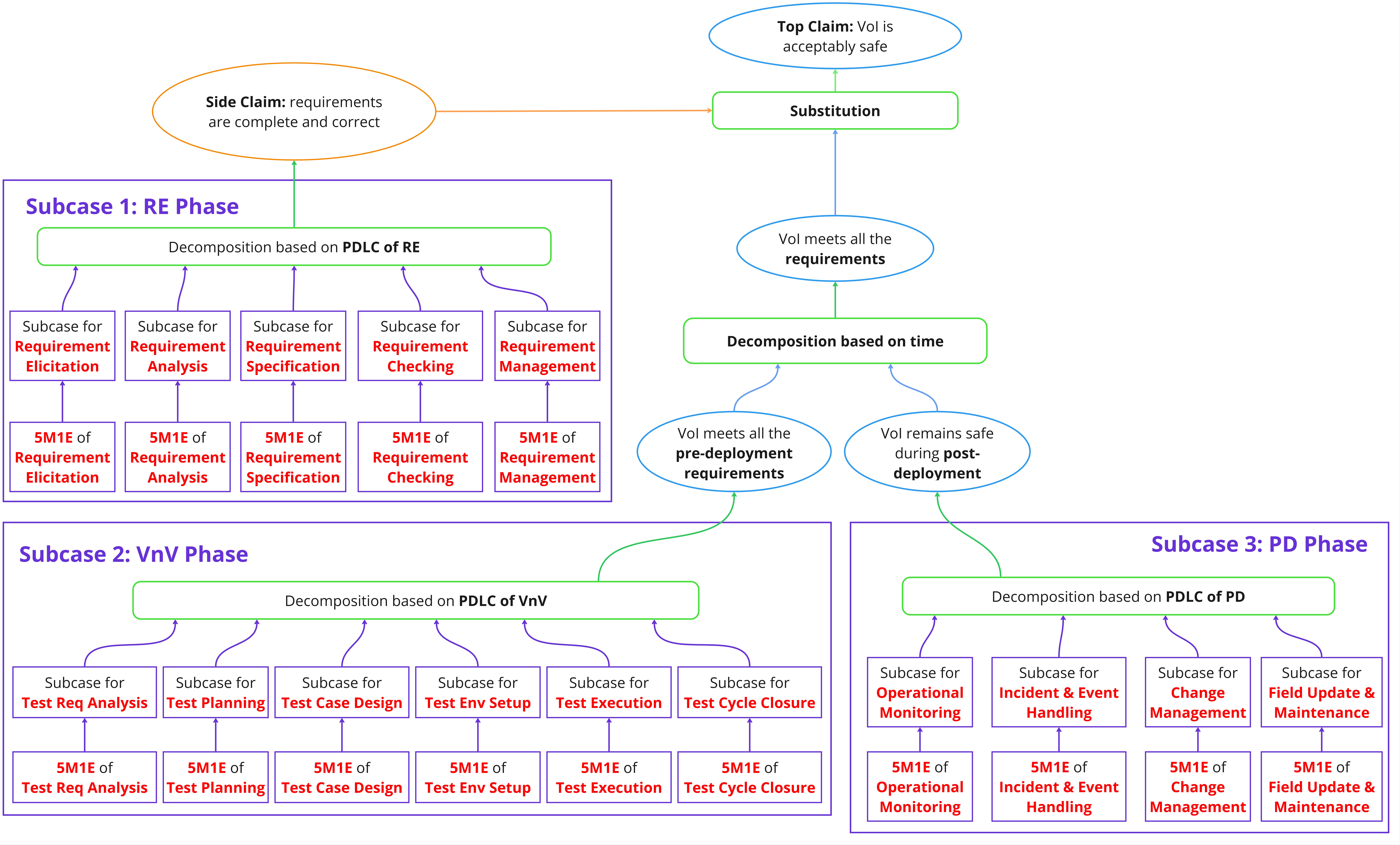}
    \caption{Level 3 Decompositions based on the 5M1E Model and their Subcases}
    \label{fig:L3 Decom}
\end{figure*}

Figure \ref{fig:L3 Decom} illustrates an overview of the proposed Level 3 decomposition based on the 5M1E model and the decomposed subcases. As a next-level decomposition of the diagram in Figure \ref{fig:L2 Decom}, each lifecycle subcase is further decomposed into the 5M1E factors of the lifecycle. 

\begin{figure*}
    \centering
    \includegraphics[width=0.95 \linewidth]{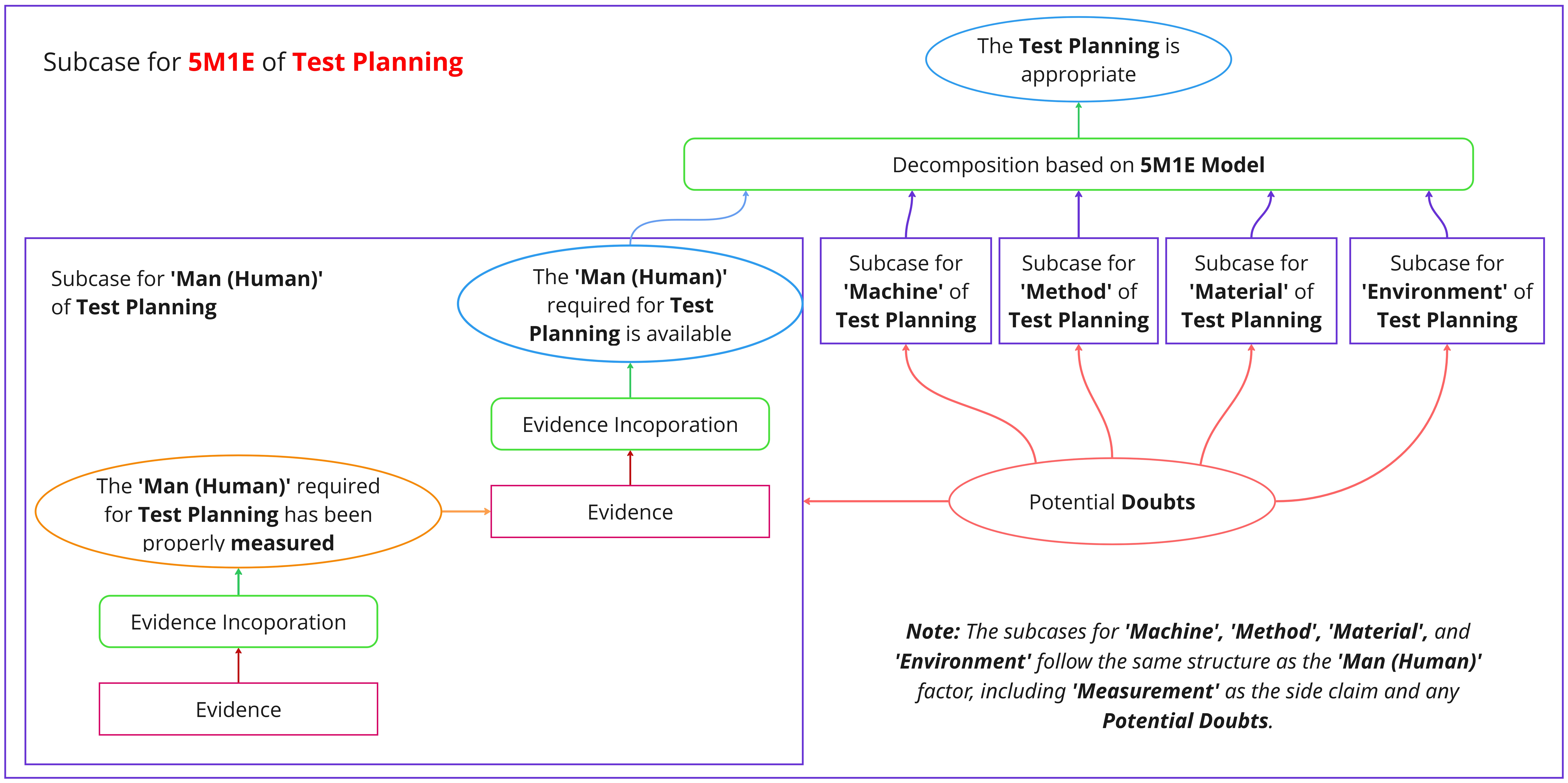}
    \caption{5M1E of Test Planning, part of the Level 3 Decompositions}
    \label{fig:Measurement}
\end{figure*}

Consider the Test Planning subcase as an example, as illustrated in Figure \ref{fig:Measurement}. The top claim of the subcase is \textbf{The Test Planning is Appropriate}. This is then further decomposed (using the Decomposition argument block) into a set of sub-subcases. This includes the sub-subcases for:
\begin{itemize}
    \item Man (Human) of Test Planning
    \item Machine of Test Planning
    \item Method of Test Planning
    \item Material of Test Planning
    \item Environment of Test Planning
\end{itemize}

Based on the new definition of the 5M1E model in Table \ref{Tab:New 5M1E}, the Measurement factor covers the measurement of the other parts of 5M1E, including the measurement of Man (Human), Method, Machine, Material, and Environment. To reflect on the definition, the Measurement part of the 5M1E is therefore presented as a side claim that points to each evidence of the sub-subcases. This ensures that the confidence level of each evidence is properly measured.

\subsection{Excel-based Templates based on the proposed Decomposition Framework}

The proposed three-level decomposition framework provides a comprehensive structure to capture all the safety arguments and evidence needed. However, an exhaustive assurance case that strictly follows the decomposition framework can end up with a minimum of 150 subcases, which include fifteen PDLC subcases (i.e., five PDLC subcases from the RE phase, six PDLC subcases from the VnV phase, and four PDLC subcases from the PD phase), each with ten subcases based on the 5M1E model. With the addition of doubts, the whole Assurance 2.0 diagram can become massive. \textcolor{black}{Creating the assurance case diagram to represent the huge amount of data without supporting tools becomes a real challenge.} To tackle this challenge, as part of the development work, a set of Excel-based worksheets was created to capture the necessary information based on the decomposition framework. A snapshot of the sample worksheet template was created as illustrated in Table \ref{Tab:Sample Template}. The template covers the Man (Human) argument and its Measurement argument of the VnV phase. Each row represents the PDLC of the VnV Phase, according to Table \ref{Tab:PDLC VnV}. The columns can be split into two groups, where the first group (i.e., columns 2 and 3) contributes to the Man (Human) argument, including the sub-claims and evidence. The second group (i.e., columns 4 and 5) contributes to the Measurement argument of Man (Human). It includes the Measurement Claim (MC) and Measurement Evidence (ME). In this work, there are in total fifteen worksheets created (i.e., Man, Machine, Material, Method, and Environment of the RE, VnV, and PD phases).



\begin{table*}
\caption{A Snapshot of the Sample Template for the Man (Human) Argument and its Measurement Argument for VnV Phase}
\label{Tab:Sample Template}
\centering

\begin{tabular}{>{\centering\arraybackslash}p{0.2\textwidth}>{\centering\arraybackslash}p{0.15\textwidth}>{\centering\arraybackslash}p{0.15\textwidth}>{\centering\arraybackslash}p{0.15\textwidth}>{\centering\arraybackslash}p{0.15\textwidth}}\toprule
\multicolumn{3}{c}{Man (Human) Argument - Decomposed based on PDLC (as each row below)}& \multicolumn{2}{c}{Measurement of Man (Human)}\\\midrule
 PDLC& Sub-claims & Evidence & Measurement Claim (MC) & Measurement Evidence (ME) \\\hline
Test Requirement Analysis &  &  &  &  \\\hline
Test Planning &  &  &  &  \\\hline
Test Case Design &  &  &  &  \\\hline
Test Environment Setup &  &  &  &  \\\hline
Test Execution &  &  &  &  \\\hline
Test Cycle Closure &  &  &  &  \\ \bottomrule

\end{tabular}
\end{table*}

\section{Case Study}
In this paper, the proposed three-level decomposition frameworks were applied in Wayve's AV2.0, which is an autonomous driving system based on E2E ML. Unlike traditional modular systems that separate perception, planning, and control, the system uses a single neural network trained directly from data to perform all driving tasks \cite{e2eai}\cite{E2E}. The following subsections present part of the assurance case for the product development of AV2.0, including the Material part of the RE Phase, the Man (Human) part of the VnV Phase, and the Method part of the PD Phase. \textcolor{black}{Whilst all the factors of 5M1E across the PDLC and phases are important, the selected ones stand out. Because the requirements (as Material of the RE phase) determine the overall development and training goals of the AI system. The human factor (Man of the VnV phase) ensures that the AI system is tested by the personnel with suitable capabilities. The procedures of continuous monitoring the AI system (Method of the PD phase) are equally important because AI models are sensitive to data drift and concept drift, and post-deployment methods like online validation, anomaly detection, and retraining protocols help support diagnostics, predictive maintenance, and system health \cite{stein2024role}.}

\subsection{Subcase 1: Material of the RE Phase and its Measurement Argument}

Table \ref{Tab:Material RE} summarises the Material and its Measurement argument during the RE phase. The Material, as part of the 5M1E in this phase, involves a diverse set of artefacts required to enable the progress of each PDLC. At the start of the RE phase, the high-level requirements related to the definition of the Operational Design Domain (ODD) and Behaviour, and customer requirements are required as evidence to argue for the availability of the artefacts for the Requirement Elicitation. Such evidence will be measured against the completeness and clarity, as well as approval from stakeholders.

Once the high-level requirements have been collected and validated, the set of requirements is then further refined and prioritised during the Requirement Analysis PDLC. The evidence required for this lifecycle includes refined and prioritised sets of requirements, together with the requirement analysis report. Each of the documented requirements is measured against a set of quality criteria suggested in ISO 29148 (Requirement Engineering), which ensures that the requirements are accurate and align with high-level requirements, consistent in formats, complete for the purpose, compliant with relevant standards or legislations, and unitary with a single, cohesive need or capability. As part of the development of an E2E AI system, the dataset used for training and testing the AI system is the key artefact. Assuring the quality of the dataset is equally important. Whilst ISO PASS 8800 outlines a set of dataset-related safety properties, the framework called OASISS (ODD-based AI Safety In autonomouS Systems) provides an evolved scoring-based mechanism for assuring the dataset used in the scenario-based training and testing of E2E systems \cite{OASISS}. To fulfil the measurement argument for the artefacts related to the Requirement Analysis PDLC, both the requirements validation report and the OASISS scoring report are required as evidence. 

It is important to note that the artefacts may overlap in several PDLCs. For example, according to \ref{Tab:PDLC RE}, the Requirement Specification PDLC involves documentation of the requirements in a structured, clear, and traceable format. The output artefacts of this PDLC and its argument are already covered in the Requirement Analysis PDLC. The Requirement Checking PDLC involves validating the documented requirements, which are covered in the measurement argument of the requirement analysis PDLC. Both requirement specification and requirement checking are therefore noted as N/A in the worksheet in Table \ref{Tab:PDLC RE}. 

The final PDLC of the RE phase is Requirement Management, involving tracking changes, maintaining traceability, and ensuring that the requirements are updated. The resultant evidence of this PDLC includes a documented requirement traceability matrix, change logs, and version history. These documents are measured against a set of criteria, including traceability, alignment with the project goals, and effectiveness of the change management. The Requirement Health Reports are documented as the evidence that summarises the requirement traceability status, requirement status, requirement coverage, their logs or change impact, and stakeholder approval status.

\begin{table*}
\centering
\caption{A Snapshot of the Excel-based Template of the Material Argument and its Measurement Argument of the RE Phase}
\label{Tab:Material RE}

\begin{tabular}{>{\centering\arraybackslash}p{0.1\textwidth}>{\raggedright\arraybackslash}p{0.15\textwidth}>{\raggedright\arraybackslash}p{0.15\textwidth}>{\raggedright\arraybackslash}p{0.22\textwidth}>{\raggedright\arraybackslash}p{0.23\textwidth}}\toprule     
\multicolumn{3}{>{\centering\arraybackslash}p{0.4\textwidth}}{Material Argument of each PDLC of RE Phase} & \multicolumn{2}{>{\centering\arraybackslash}p{0.45\textwidth}}{Measurement of Material} \\\midrule

PDLC & Sub-claims & Evidence & Measurement Claim (MC) & Measurement Evidence (ME) \\\hline
Requirement Elicitation & The artefacts for the requirement elicitation cycle are appropriate & E1: Documented High-level requirements for the ODD Definition & MC1: The documented High-level requirements for the ODD Definition has met the measurement criteria are complete and correct & ME1: Documented initial requirement review report \\
&  & E2: Documented High-level requirements for the Behaviour Definition & MC2: The documented High-level requirements for the Behaviour Definition are complete and correct & ME2: Documented initial requirement review report \\
&  & E3: Documented Customer Requirements & MC3: The documented Customer Requirements has met the stakeholder satisfaction & ME3: Documented stakeholder satisfaction report \\\hline

Requirement Analysis & The artefacts for the requirement analysis cycle are appropriate & E1: Documented Refined sets of requirements (functional, non-functional) & MC1: The documented Refined sets of requirements (functional, non-functional) has met the requirement quality criteria suggested in relevant standards (ISO 29148, IEEE 830, ISO 8800) & ME1: Documented Requirements Validation Report \\
 &  & E2: Documented Prioritized requirements & MC2: The documented Prioritized requirements has met the requirement quality criteria suggested in relevant standards (ISO 29148, IEEE 830, ISO 8800, OASISS) & ME2: Documented Requirements Validation and OASISS Scoring Report \\
 &  & E3: Documented Requirement analysis report & MC3: The documented Requirement analysis report has passed the audit check & ME3: Audit check report \\\hline
Requirement Specification & \multicolumn{2}{>{\centering\arraybackslash}p{0.3\textwidth}}{N/A - covered in Requirement Analysis} & \multicolumn{2}{>{\centering\arraybackslash}p{0.45\textwidth}}{N/A - covered in Measurement of Requirement Analysis} \\\hline

Requirement Checking & \multicolumn{2}{>{\centering\arraybackslash}p{0.3\textwidth}}{N/A - covered in Measurement of Requirement Analysis} & \multicolumn{2}{>{\centering\arraybackslash}p{0.45\textwidth}}{N/A - covered in Measurement of Requirement Analysis} \\\hline

Requirement Management & The artefacts for the requirement management cycle are appropriate & E1: Documented Requirement traceability matrix & MC1: The documented Requirement traceability matrix has passed the audit check & ME1: Audit check report against traceability and alignment with project goals \\

 &  & E2: Documented Change logs & MC2: The documented Change logs has passed the audit check & ME2: Audit check report \\
 &  & E3: Documented Version history & MC3: The documented Version history has passed the audit check & ME3: Audit check report \\ \bottomrule

\end{tabular}
\end{table*}

\subsection{Subcase 2: Man (Human) of the VnV Phase and its Measurement Argument}

Table \ref{Tab:Man VnV} summarizes the collected information related to the Man (Human) and its corresponding Measurement argument during the VnV phase. This phase involves a wide range of roles and responsibilities to support the claims associated with each stage of the PDLC.

During the initial Test Requirement Analysis PDLC, the test manager, test lead, and test analysts assess the feasibility of testing and identify testable requirements. Developers contribute technical insights from various subsystems, including the braking system, powertrain, and AI components. 

In the Test Planning PDLC, the test manager develops detailed test plans, while the project manager ensures alignment with the overall project timeline. The QA lead also participates in estimating the required effort and resources.

Moving on to the Test Case Design PDLC, test analysts create test cases with input from developers, who provide technical data. These test cases are then reviewed and approved by the test lead. 

As the product development proceeds to the Test Environment Setup PDLC, system administrators and developers configure the test environment. The test manager and testers validate the setup to ensure readiness. 

Once the test cases and environments are finalized, test engineers execute the tests and log any defects encountered. These defects are reported back to the developers for resolution. The test lead oversees the entire test execution process to ensure consistency and completeness. 

Finally, during the Test Cycle Closure PDLC, the test manager compiles a comprehensive test summary report, and the QA lead contributes to the retrospective analysis to evaluate the overall testing process and outcomes.

The availability of relevant human resources across the PDLC ensures that the VnV phase is equipped with the right Man (Human) element. This is supported by the resumes of each role as evidence of the sub-claims. As part of the assurance case development, each part of the evidence needs to be validated via the Measurement argument. In the same table (i.e., Table \ref{Tab:Man VnV}), each evidence is assessed as presented in the Measurement Claim and Measurement Evidence columns. In this work, the Suitably Qualified and Experienced Person (SQEP) is referred to as the criterion to measure the quality of the resumes \cite{sqep}. A SQEP is someone who possesses the appropriate qualifications, experience, and competence to perform specific tasks or roles safely and effectively. To be considered a SQEP, the individual typically must demonstrate relevant qualifications, practical experience, up-to-date knowledge, analytical skills, and communication skills. For each role, the detailed content of each SQEP criterion can be different. For example, consider the SQEP criteria for a test manager in the context of the product development of an E2E AI system. To fulfil the Qualification criterion, the individual must have a relevant degree in computer science, software engineering, AI, or a related field. The test manager must also have a proven track record in managing test activities for complex, multi-component systems, with experience in AI/ML model validation, especially in real-time or embedded environments (i.e., the Experience criterion of SQEP). The test manager must also have a good understanding of E2E AI architectures and knowledge of relevant simulation tools and test automation frameworks (i.e., the Knowledge criterion of SQEP), with competence in VnV planning, execution, reporting, risk-based testing, and safety assurance practices (i.e., the Technical Competence criterion of SQEP). Furthermore, the ideal test manager must also have the ability to lead cross-functional teams, with effective stakeholder communication skills and commitment to staying updated with AI safety standards, testing tools, and industry trends (i.e., the Soft Skills criterion of SQEP). The SQEP criteria for each of the roles are assessed, either via interviews or workshops, and documented in the final SQEP Check Report as the Measurement Evidence.

\begin{table*}
\centering
\caption{A Snapshot of the Excel-based Template of the Man (Human) Argument and its Measurement Argument of the VnV Phase}
\label{Tab:Man VnV}

\begin{tabular}{>{\centering\arraybackslash}p{0.1\textwidth}>{\raggedright\arraybackslash}p{0.15\textwidth}>{\raggedright\arraybackslash}p{0.12\textwidth}>{\raggedright\arraybackslash}p{0.25\textwidth}>{\raggedright\arraybackslash}p{0.23\textwidth}}\toprule     
\multicolumn{3}{>{\centering\arraybackslash}p{0.37\textwidth}}{Man (Human) Argument of each PDLC of VnV Phase} & \multicolumn{2}{>{\centering\arraybackslash}p{0.48\textwidth}}{Measurement Argument of Man (Human)} \\\midrule
PDLC & Sub-claims & Evidence & Measurement Claim (MC) & Measurement Evidence (ME) \\\hline
Test Requirement Analysis & The personnel responsible for Test Requirement Analysis are appropriate & E1: Resume for the Test Manager & MC1: The Test Manager has met the SQEP (Suitably Qualified Experienced Personnel) Criteria & ME1: SQEP Check Report of the Test Manager \\
 &  & E2: Resume for the Test Lead & MC2: The Test Lead has met the SQEP (Suitably Qualified Experienced Personnel) Criteria & ME2: SQEP Check Report of the Test Lead \\
 &  & E3: Resume for the Developers & MC3: The Developers has met the SQEP (Suitably Qualified Experienced Personnel) Criteria & ME3: SQEP Check Report of the Developers \\\hline
Test Planning & The personnel responsible for Test Planning are appropriate & E1: Resume for the Test Manager & MC1: The Test Manager has met the SQEP (Suitably Qualified Experienced Personnel) Criteria & ME1: SQEP Check Report of the Test Manager \\
 &  & E2: Resume for the Project Manager & MC2: The Project Manager has met the SQEP (Suitably Qualified Experienced Personnel) Criteria & ME2: SQEP Check Report of the Project Manager \\
 &  & E3: Resume for the QA Lead & MC3: The QA Lead has met the SQEP (Suitably Qualified Experienced Personnel) Criteria & ME3: SQEP Check Report of the QA Lead \\\hline
Test Case Design & The personnel responsible for Test Case Design are appropriate & E1: Resume for the Test Analyst & MC1: The Test Analyst has met the SQEP (Suitably Qualified Experienced Personnel) Criteria & ME1: SQEP Check Report of the Test Analyst \\
 &  & E2: Resume for the Test Lead & MC2: The Test Lead has met the SQEP (Suitably Qualified Experienced Personnel) Criteria & ME2: SQEP Check Report of the Test Lead \\
 &  & E3: Resume for the Developers & MC3: The Developers has met the SQEP (Suitably Qualified Experienced Personnel) Criteria & ME3: SQEP Check Report of the Developers \\\hline
Test Environment Setup & The personnel responsible for Test Environment Setup are appropriate & E1: Resume for the System Administrator & MC1: The System Administrator has met the SQEP (Suitably Qualified Experienced Personnel) Criteria & ME1: SQEP Check Report of the System Administrator \\
 &  & E2: Resume for the Test Manager & MC2: The Test Manager has met the SQEP (Suitably Qualified Experienced Personnel) Criteria & ME2: SQEP Check Report of the Test Manager \\
 &  & E3: Resume for the Testers & MC3: The Testers has met the SQEP (Suitably Qualified Experienced Personnel) Criteria & ME3: SQEP Check Report of the Testers \\\hline
Test Execution & The personnel responsible for Test Environment Setup are appropriate & E1:Resume for the Test Engineers & MC1: The Test Engineers has met the SQEP (Suitably Qualified Experienced Personnel) Criteria & ME1: SQEP Check Report of the Test Engineers \\
 &  & E2: Resume for the Test Lead & MC2: The Test Lead has met the SQEP (Suitably Qualified Experienced Personnel) Criteria & ME2: SQEP Check Report of the Test Lead \\
 &  & E3: Resume for the Developers & MC3: The Developers has met the SQEP (Suitably Qualified Experienced Personnel) Criteria & ME3: SQEP Check Report of the Developers \\\hline
Test Cycle Closure & The personnel responsible for Test Cycle Closure are appropriate & E1: Resume for the Test Manager & MC1: The Test Manager has met the SQEP (Suitably Qualified Experienced Personnel) Criteria & ME1: SQEP Check Report of the Test Manager \\
 &  & E2: Resume for the QA Lead & MC2: The QA Lead has met the SQEP (Suitably Qualified Experienced Personnel) Criteria & ME2: SQEP Check Report of the QA Lead \\ \bottomrule

\end{tabular}
\end{table*}

\subsection{Subcase 3: Method of the PD Phase and its Measurement Argument}

Table \ref{Tab:Method PD} summarises the Method and its Measurement argument for the PD phase.

During the initial Operational Monitoring PDLC, both the real-world conditions and cybersecurity of the deployed vehicle are monitored. These are achieved via continuous monitoring and logging based on real-time data collection from sensors and systems, fleet learning by aggregating data across vehicles, statistical process control by monitoring system performance metrics over time, threat analysis and risk assessment (TARA), and collection of shared data from other ADS and infrastructure.

If the deployed vehicle is involved in any incident or accident, the next PDLC - i.e., Incident and Event Handling will be implemented. This PDLC involves incident analysis techniques such as Causal Analysis based on System Theory (CAST) \cite{CAST}, System Theoretic Process Analysis (STPA) \cite{STPA1} \cite{STPA2}, and AcciMap \cite{AcciMap}, etc. Based on the analysis results, corrective and preventive actions are identified to provide structured responses to the incidents.

Based on the outputs from the previous PDLCs, corresponding changes to the vehicle system are identified, assessed, implemented, and validated in the Change Management PDLC. The applied methods for change management include:

\begin{itemize}
    \item HARA, functional decomposition, and use case modelling for impact assessment;
    \item  In-the-loop testing and scenario-based testing for re-verification and re-validation (re-VnV);
    \item Decomposition and management of updated requirements;
    \item Updating Assurance 2.0 to align with the change;
    \item Joint live documents and dedicated meetings for stakeholder communication;
\end{itemize}

The last PDLC of the PD phase - i.e., Field Update and Maintenance, includes:

\begin{itemize}
    \item Updates and maintenance of the system. This includes DevOps Pipelines for continuous integration and deployment of software updates;
    \item Change impact analysis to assess the effect of updates on system safety;
    \item Security Patch Management to provide regular updates to address known threats;
    \item All of the evidence is reviewed and audited for compliance with relevant standards, including ISO/TS 5083 and ISO 21434 \cite{ISO21434}.
\end{itemize}

\begin{table*}
\centering
\caption{A Snapshot of the Excel-based Template of the Method Argument and its Measurement Argument of the PD Phase}
\label{Tab:Method PD}

\begin{tabular}{>{\centering\arraybackslash}p{0.1\textwidth}>{\raggedright\arraybackslash}p{0.12\textwidth}>{\raggedright\arraybackslash}p{0.18\textwidth}>{\raggedright\arraybackslash}p{0.22\textwidth}>{\raggedright\arraybackslash}p{0.23\textwidth}}\toprule     
\multicolumn{3}{>{\centering\arraybackslash}p{0.4\textwidth}}{Method Argument of each PDLC of PD Phase} & \multicolumn{2}{>{\centering\arraybackslash}p{0.45\textwidth}}{Measurement of Method} \\\midrule

PDLC & Sub-claims & Evidence & Measurement Claim (MC) & Measurement Evidence (ME) \\\hline
Operational Monitoring & The process or technique for the Operational Monitoring is appropriate & E1: Documented techniques re. continuous monitoring and logging & MC1: The documented techniques re. continuous monitoring and logging has passed the audit check & ME1: Audit report based on ISO/TS 5083 and ISO 21434 \\
 &  & E2: Documented techniques re. fleet learning & MC2: The documented techniques re. fleet learning has passed the audit check & ME2: Audit report based on ISO/TS 5083 and ISO 21434 \\
 &  & E3: Documented techniques re. statistical process control & MC3: The documented techniques re. statistical process control has passed the audit check & ME3: Audit report based on ISO/TS 5083 and ISO 21434 \\
 &  & E4: Documented techniques re. Threat Analysis and Risk Assessment (TARA) & MC4: The documented techniques re. Threat Analysis and Risk Assessment (TARA) has passed the audit check & ME4: Audit report based on ISO/TS 5083 and ISO 21434 \\
 &  & E5: Documented techniques re. collection of shared data from other ADS and infrastructure & MC5: The documented techniques re. collection of shared data from other ADS and infrastructure has passed the audit check & ME5: Audit report based on ISO/TS 5083 and ISO 21434 \\\hline
Incident and Event Handling & The process or technique for Incident and Event Handling is appropriate & E1: Documented techniques re. incident analysis & MC1: The documented techniques re. incident analysis has passed the audit check & ME1: Audit report based on ISO/TS 5083 and ISO 21434 \\
 &  & E2: Documented techniques re. corrective and preventive actions & MC2: The documented techniques re. corrective and preventive actions has passed the audit check & ME2: Audit report based on ISO/TS 5083 and ISO 21434 \\\hline
Change Management & The process or technique used for Change Management is appropriate & E1: Documented techniques re. impact assessment & MC1: The documented process re. impact assessment has passed the audit check & ME1: Audit report based on ISO/TS 5083 and ISO 21434 \\
 &  & E2: Documented techniques re. system re-VnV & MC2: The documented process re. system re-VnV has passed the audit check & ME2: Audit report based on ISO/TS 5083 and ISO 21434 \\
 &  & E3: Documented techniques re. configuration management & MC3: The documented process re. configuration management has passed the audit check & ME3: Audit report based on ISO/TS 5083 and ISO 21434 \\
 &  & E4: Documented techniques re. safety case update & MC4: The documented process re. safety case update has passed the audit check & ME4: Audit report based on ISO/TS 5083 and ISO 21434 \\
 &  & E5: Documented techniques re. stakeholder communication & MC5: The documented process re. stakeholder communication has passed the audit check & ME5: Audit report based on ISO/TS 5083 and ISO 21434 \\\hline
Field Update and Maintenance & The process or technique for Field Update and Maintenance is appropriate & E1: Documented process based on DevOps Pipelines & MC1: The documented process based on DevOps Pipeline has passed the audit check & ME1: Audit report based on ISO/TS 5083 and ISO 21434 \\
 &  & E2: Documented process based on Change Impact Analysis & MC2: The documented process based on Change Impact Analysis has passed the audit check & ME2: Audit report based on ISO/TS 5083 and ISO 21434 \\
 &  & E3: Documented process based on Security Patch Management & MC3: The documented process based on Security Patch Management has passed the audit check & ME3: Audit report based on ISO/TS 5083 and ISO 21434 \\ \bottomrule

\end{tabular}
\end{table*}

\section{Discussion}

In this paper, a scalable and comprehensive assurance framework for Self-Driving Vehicles (SDVs) is presented as an extension of the existing Assurance 2.0 paradigm. The proposed approach introduces a three-level decomposition strategy designed to address the increasing complexity, adaptiveness, and autonomy of modern systems, particularly those integrating E2E AI/ML.

At Level 1, the framework decomposes vehicle-level safety assurance according to the V-model. This ensures the end-to-end coverage of safety arguments throughout the system lifecycle, from the initial design phase (Requirements Engineering, RE), through Verification and Validation (VnV), to Post-Deployment (PD). The RE phase focuses on defining system requirements, the VnV phase ensures the system is correctly built and meets its intended purpose, and the PD phase monitors the system in real-world conditions to ensure ongoing safety, security, and effectiveness.

At Level 2, each of the three lifecycle phases is further decomposed using Product Development Lifecycle (PDLC) models derived from internationally recognized standards. Specifically, the RE phase is based on ISO 29148 (Requirements Engineering), the VnV phase on ISO 29119 (Software Testing), and the PD phase on ISO/TS 5083, which emphasizes post-deployment monitoring as a critical component of safety assurance for Automated Driving Systems (ADS). This standards-based decomposition ensures both regulatory alignment and domain independence.

At Level 3, each PDLC is further decomposed using an adapted version of the 5M1E model—Man, Machine, Method, Material, Measurement, and Environment. Unlike the traditional manufacturing-focused 5M1E model, each element is redefined to suit the assurance domain and structured to be mutually exclusive yet interconnected. This interrelationship is illustrated in the onion diagram (Figure \ref{fig:Onion Diagram}), where the outermost layer, Measurement, serves as a meta-level factor that evaluates the confidence in all other elements.

To manage the complexity of the resulting assurance case, which may involve over 150 subcases, Excel-based templates were developed to support structured documentation and automation. These templates serve as inputs to toolchains that generate Assurance 2.0 diagrams, enabling continuous and incremental assurance.

A case study involving Wayve’s AV2.0 system, an E2E AI-based autonomous driving platform, demonstrates the practical application of the framework. The decomposition was applied to various assurance aspects, including the quality of training datasets (Material and its Measurement in RE, Table \ref{Tab:Material RE}), personnel qualifications (Man and its Measurement in VnV, Table \ref{Tab:Man VnV}), and post-deployment monitoring methods (Method and its Measurement in PD, Table \ref{Tab:Method PD}). These examples highlight the framework’s scalability and rigor.


The proposed decomposition frameworks directly respond to the three research questions outlined in the paper. The multi-level decomposition ensures exhaustive coverage of safety arguments across lifecycle stages and operational dimensions. By integrating internationally recognised standards and the 5M1E model, the framework captures all relevant safety concerns, evidence, and counterarguments. This addresses the first research question - i.e., \textbf{RQ1: Completeness}. The use of structured templates and decomposition logic supports automated generation of the final Assurance 2.0 diagrams. This reduces manual effort, minimizes inconsistencies, and enables agile adaptation to system changes. This addresses the second research question - i.e., \textbf{RQ2: Efficiency}. Finally, the hierarchical structure and standardized decomposition improve clarity and traceability. Stakeholders from diverse disciplines can navigate the assurance case more easily, enhancing cross-functional understanding and regulatory engagement. This addresses the last research question - i.e., \textbf{RQ3: Communication}.


Unlike previous templates tailored to specific machine learning assurance scenarios, the integration of international standards enables scalable applicability across abstraction levels. For instance, a Tier-1 supplier specializing in automated driving solutions can start the safe case development with the ADS-level top claim. Followed by the application of  the framework to decompose the top claim into RE, VnV, and PD phases of the ADS development, followed by PDLC and 5M1E decompositions. The three-level framework can also be universal to other domains. For example, considering applying the decomposition frameworks for the safety case of an aircraft system, the top-claim (i.e., the aircraft level claim) can be decomposed following the suggested phases and lifecycles in ARP4761 and ARP4754A \cite{ARP4761} \cite{ARP4754A_2010}, followed by the level-3 decomposition based on the 5M1E model.
\section{Conclusion}
A scalable and universal framework based on Assurance 2.0 to efficiently and effectively identify and manage safety arguments was presented in this paper. The decomposition frameworks utilize the internationally recognized standards and models. An Excel-based template was also proposed in the paper to support the documentation of the safety arguments. As part of future work, a toolchain will be developed to enable the auto-generation of Assurance 2.0 diagrams by importing and extracting the information from the templates. The proposed framework integrates the 5M1E model, providing a comprehensive set of factors that need to be considered for safety assurance. The Measurement factor, based on the updated definition in Table \ref{Tab:New 5M1E}, ensures that all the evidence is properly assessed and measured for quality, reliability, and safety. The current measurement mechanisms are based on a set of qualitative measurement criteria, which are insufficient to support the argument of complex evidence. Whilst OASISS provides a comprehensive framework that contributes to the quantitative measurement of training and testing datasets for E2E AI systems, as part of future work, additional quantitative models need to be explored or developed to quantify the confidence level of evidence.

\bibliographystyle{ieeetr}
\bibliography{References}

\begin{IEEEbiography}[{\includegraphics[width=1in,height=1.25in,clip,keepaspectratio]{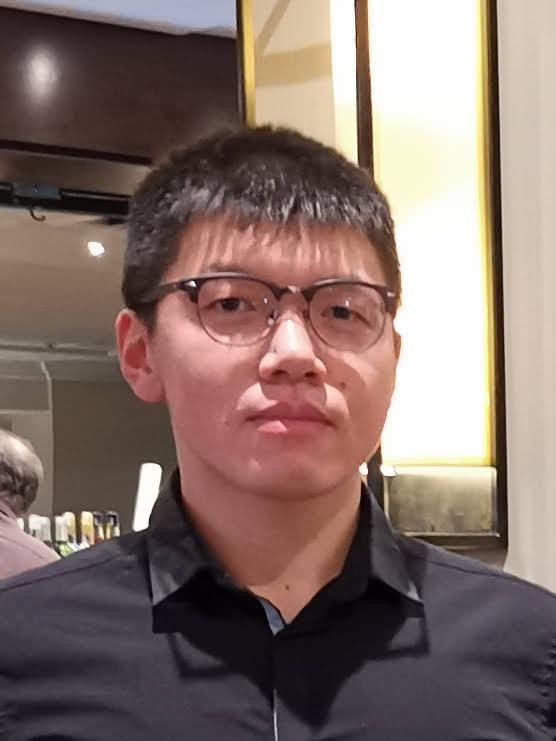}}]{Shufeng Chen}
received the B.Eng. (Hons) degree in Electronic Engineering from the University of Southampton, UK. He is currently a Lead Engineer in System Safety at WMG, University of Warwick, where he specializes in system safety analysis and safety case development for complex and emerging technologies. His research interests lie in the safety assurance of autonomous systems, end-to-end AI pipelines, aviation operations, and socio-technical integration. He has contributed to several interdisciplinary projects involving the development of safety frameworks for AI-enabled systems and has worked closely with industry and academic partners to advance safety engineering practices in real-world applications.
\end{IEEEbiography}

\begin{IEEEbiography}[{\includegraphics[width=1in,height=1.25in,clip,keepaspectratio]{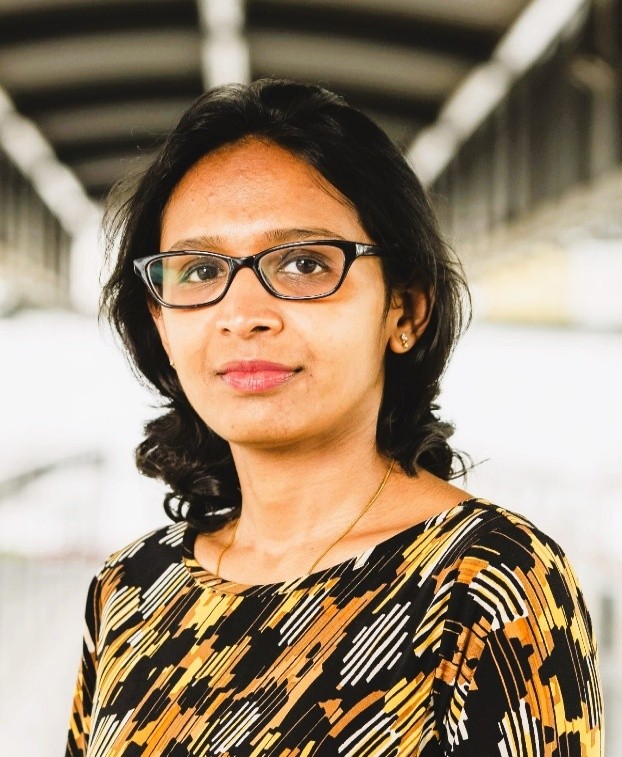}}]{MARIAT JAMES ELIZEBETH} received B.Eng. Hons in Bio-Medical Engineering from Bharath Institute of Higher Education and Research, India. She worked as a Software Engineer for Robert Bosch, India, developing In-Vehicle Infotainment Systems before joining Jaguar Land Rover, Gaydon, UK, as a system and software architect for the next-generation infotainment system development program. In 2019, she joined Warwick Manufacturing Group (WMG), University of Warwick, as a Lead Engineer in the Intelligent Vehicles group. Her skills and research interests include Requirements Engineering, System Architecture and Design, Systems Engineering, System Safety, and Autonomous Vehicles.
\end{IEEEbiography}

\begin{IEEEbiography}[{\includegraphics[width=1in,height=1.25in,clip,keepaspectratio]{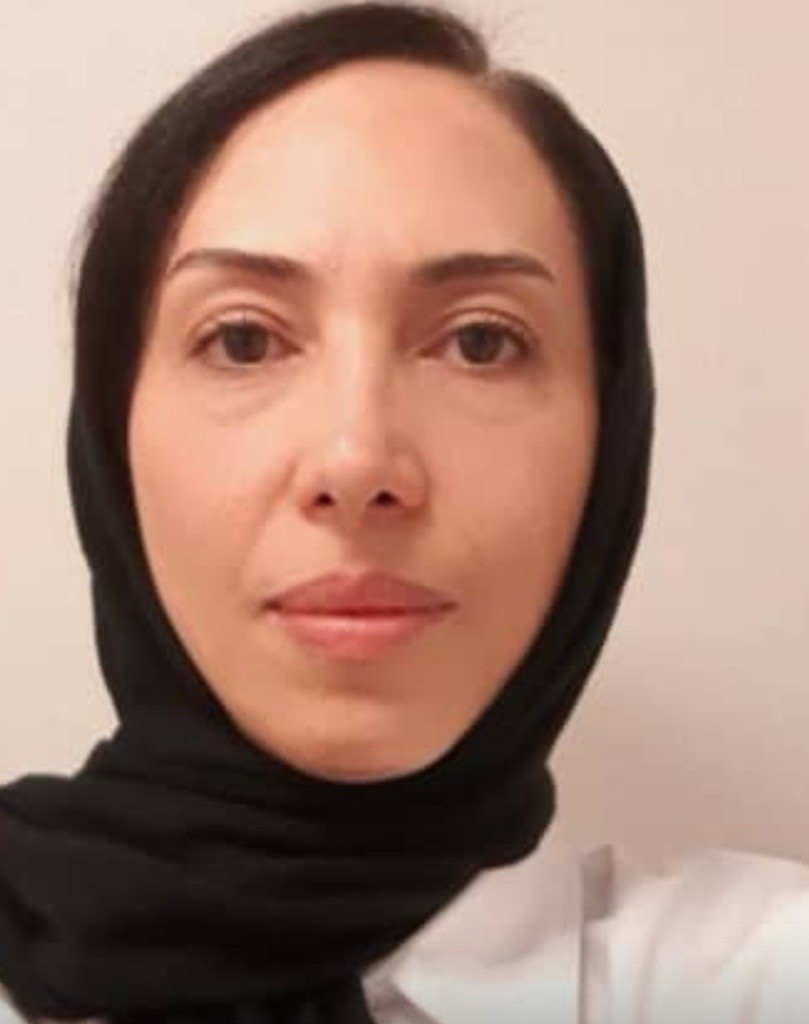}}]{Robab Aghazadeh Chakherlou}  is a Research Fellow in Safe AI within the Safe Autonomy Research Group at the University of Warwick. Her research focuses on the dependability assessment of safety-critical systems, with a particular emphasis on ensuring the reliability and robustness of AI-driven autonomous technologies. Dr. Aghazadeh Chakherlou contributes to advancing methodologies that support the safe deployment of AI in complex, real-world environments. She actively collaborates on interdisciplinary projects aimed at enhancing the trustworthiness of intelligent systems.
\end{IEEEbiography}

\begin{IEEEbiography}[{\includegraphics[width=1in,height=1.25in,clip,keepaspectratio]{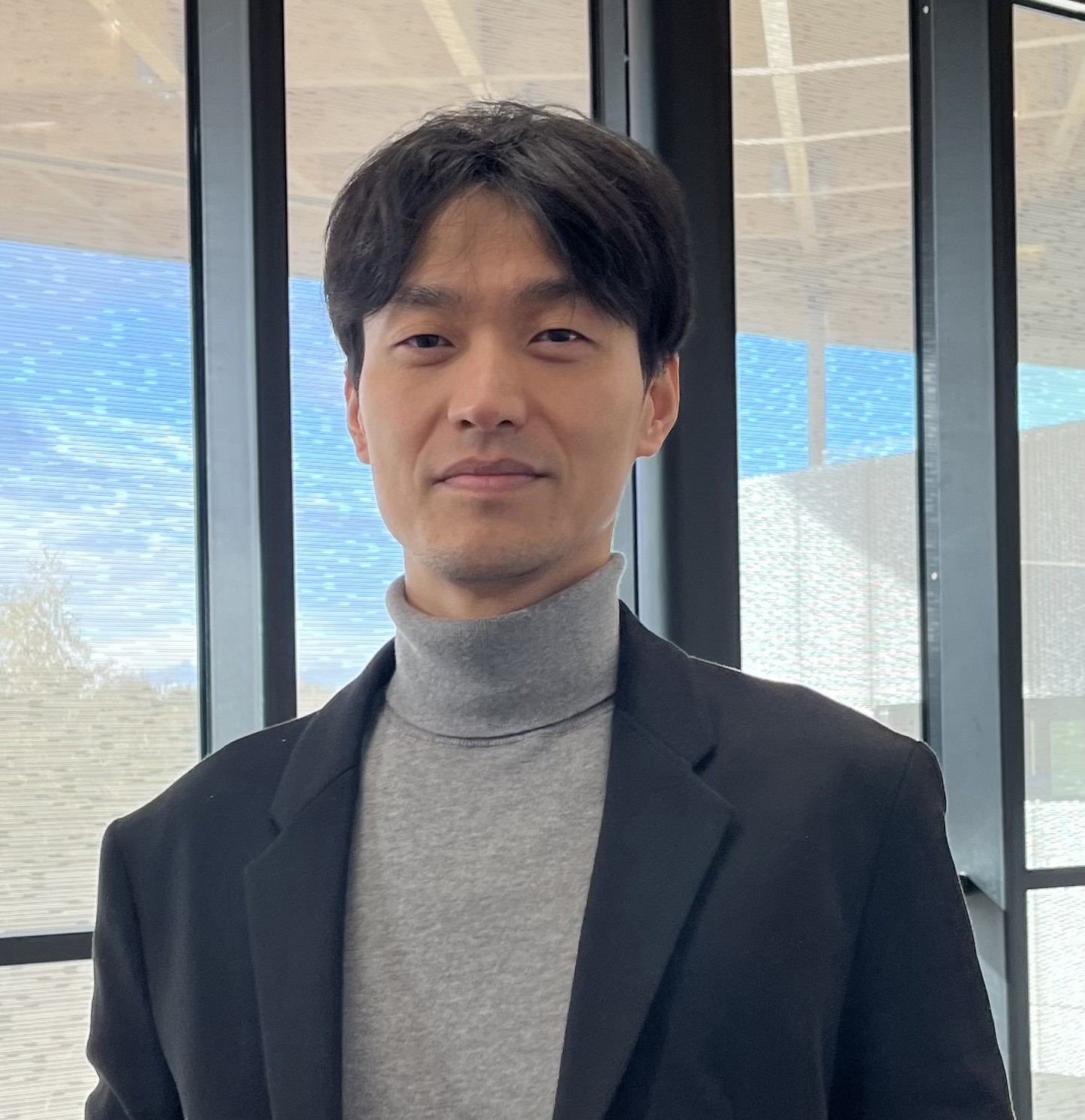}}]{Xingyu Zhao} is an Associate Professor in Safety-Critical AI Systems at WMG, University of Warwick, focusing on trust in AI for high-stakes applications like autonomous vehicles. His research spans safety assurance, formal verification, probabilistic reasoning, and adversarial robustness to enable safe, explainable AI decision-making under uncertainty. In total, he has produced 60+ publications in top-tier journals and conferences. His award-winning work includes Siemens AI-Dependability Assessment Student Challenge, UK-US Privacy-Enhancing Technologies Prize Challenges, and best papers in top conferences (AISafety@IJCAI and ISSRE).
\end{IEEEbiography}

\begin{IEEEbiography}[{\includegraphics[width=1in,height=1.25in,clip,keepaspectratio]{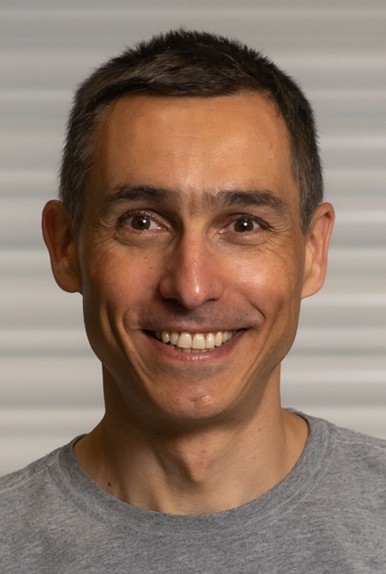}}]{Eric Barbier} is a Functional Safety expert with more than 25 years of experience in the automotive industry. He started his career as a software developer and soon shifted his focus to functional safety. He worked as an engineering consultant for thirteen years helping Tier Ones and OEMs release to market safe electric vehicles and subsystems. Eric validated his skills by gaining a postgraduate degree in Safety-Critical System Engineering from the University of York in 2017. He took on the new challenge of AI’s safety assurance when joining Wayve in 2019 and is thrilled by the technology's complexity and its tremendous potential.
\end{IEEEbiography}

\begin{IEEEbiography}[{\includegraphics[width=1in,height=1.25in,clip,keepaspectratio]{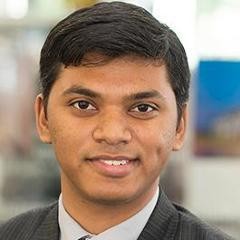}}]{Siddartha Khastgir} is a professor at WMG, University of Warwick. He holds a PhD in the area of ADS testing and Human Factors (trust in automation). He is an expert in ADS/ADAS safety verification and has significant experience in the ADS domain include test scenarios generation, safety, simulation-based testing, standardization and international regulations. He has over 10 years’ experience in safety verification and validation.
\end{IEEEbiography}

\begin{IEEEbiography}[{\includegraphics[width=1in,height=1.25in,clip,keepaspectratio]{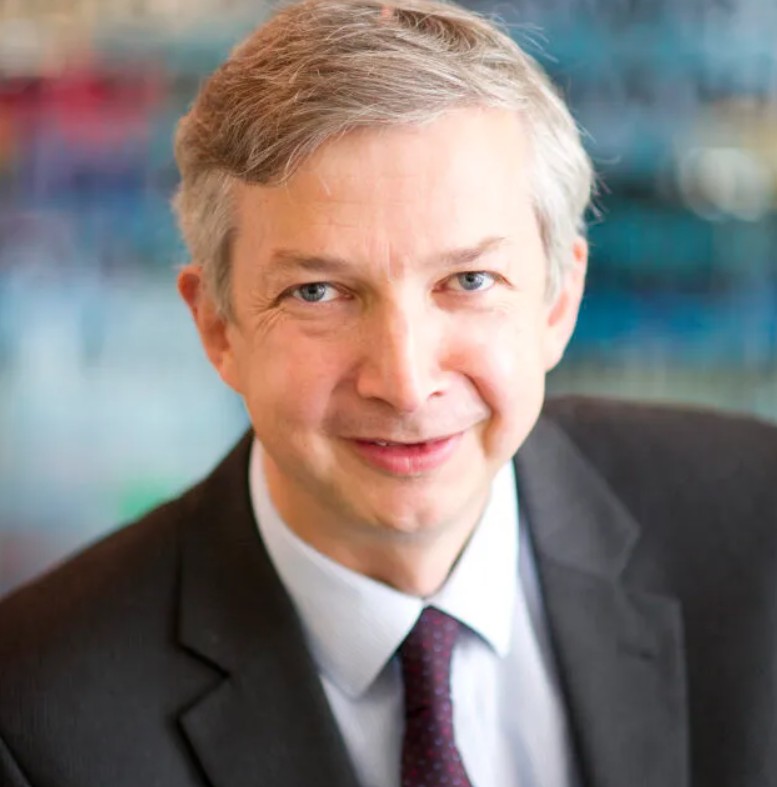}}]{Paul Jennings} is a professor and the research director at WMG, University of Warwick. He has over 30 years of experience in industrial collaborative R\&D projects, having led research projects worth over £60M. Before taking up his role as Research Director at WMG, he set up and led the Intelligent Vehicles group at WMG with a key focus on Safety. He has published over 100 academic papers and has been Principal Investigator on over 30 research grants.
\end{IEEEbiography}
\EOD

\end{document}